\documentstyle[12pt]{article}
\begin{document}

\tolerance=5000
\def\pp{{\, \mid \hskip -1.5mm =}}
\def\cL{{\cal L}}
\def\be{\begin{equation}}
\def\ee{\end{equation}}
\def\bea{\begin{eqnarray}}
\def\eea{\end{eqnarray}}
\def\tr{{\rm tr}\, }
\def\nn{\nonumber \\}
\def\e{{\rm e}}
\def\D{{D \hskip -3mm /\,}}

\def\SEH{S_{\rm EH}}
\def\SGH{S_{\rm GH}}
\def\AdS5{{{\rm AdS}_5}}
\def\S4{{{\rm S}_4}}
\def\gfv{{g_{(5)}}}
\def\gfr{{g_{(4)}}}
\def\SC{{S_{\rm C}}}
\def\RH{{R_{\rm H}}}

\def\wlBox{\mbox{
\raisebox{0.1cm}{$\widetilde{\mbox{\raisebox{-0.1cm}\fbox{\ }}}$}}}
\def\htBox{\mbox{
\raisebox{0.1cm}{$\hat{\mbox{\raisebox{-0.1cm}{$\Box$}}}$}}}

\def\K{\left(k - (d-2) \e^{-2\lambda}\right)}

\  \hfill
\begin{minipage}{3.5cm}
UPR 970-T \\
December 2001 \\
\end{minipage}

\vfill

\begin{center}
{\large\bf  Black Hole Thermodynamics and Negative Entropy in deSitter and Anti-deSitter Einstein-Gauss-Bonnet gravity }

\vfill

{\sc Mirjam CVETI\v C}\footnote{cvetic@cvetic.hep.upenn.edu},\\ 
{\sc Shin'ichi NOJIRI}$^\clubsuit$\footnote{nojiri@cc.nda.ac.jp},\\
and {\sc Sergei D. ODINTSOV}$^{\spadesuit}$\footnote{
odintsov@ifug5.ugto.mx, odintsov@mail.tomsknet.ru}\\

\vfill

{\sl Department of Physics and Astronomy, \\
University of Pennsylvania, Philadelphia PA 19104-6396, USA}

\vfill

{\sl $\clubsuit$ Department of Applied Physics \\
National Defence Academy,
Hashirimizu Yokosuka 239-8686, JAPAN}

\vfill

{\sl $\spadesuit$
Tomsk State Pedagogical University, 634041 Tomsk, RUSSIA \\
and \\
Instituto de Fisica de la Universidad de Guanajuato,
Lomas del Bosque 103, Apdo. Postal E-143, 
37150 Leon,Gto., MEXICO 
}

\vfill
\newpage

{\bf ABSTRACT}

\end{center}
We investigate the charged Schwarzschild-Anti-deSitter (SAdS) BH 
thermodynamics in 5d Einstein-Gauss-Bonnet 
gravity with electromagnetic field. The Hawking-Page phase 
transitions between SAdS BH and pure AdS space are studied.
The corresponding phase diagrams (with critical line defined by GB term 
coefficient and electric charge) are drawn. The possibility to account 
for higher derivative Maxwell terms is mentioned. 

In frames of proposed dS/CFT correspondence it is demonstrated that
brane gravity maybe localized similarly to AdS/CFT. SdS BH 
thermodynamics in 5d Einstein and Einstein-Gauss-Bonnet gravity is
 considered.
 The corresponding (complicated) surface 
counterterms are found and used to get the conserved BH mass, free energy 
and entropy.
The interesting feature of higher derivative gravity is the possibility 
for negative (or zero) SdS (or SAdS) BH entropy which  depends on the 
parameters of higher derivative terms. We speculate that the appearence 
of negative entropy may indicate a  
 new type instability where a transition between
SdS (SAdS) BH with negative entropy to SAdS (SdS) BH with positive 
entropy  would occur.


\newpage

\section{Introduction}

The holographic principle in 
M-theory/superstrings physics in the form of AdS/CFT 
correspondence \cite{AdS} and recently dS/CFT
correspondence\cite{strominger,ds1} is one of the main motivations to study 
 the black hole thermodynamics with asymptotic Anti-deSitter (AdS) and 
deSitter (dS) spacetimes. Indeed, such classical gravity black holes turned 
out to describe the properties of dual quantum field 
theory living on the boundary of corresponding black hole (BH). 
Hence, two, so far distant branches of theoretical physics appear as
different manifestations of the same theory.

There are different approaches to describe BH thermodynamics.
Even the starting theory maybe different. Traditionally, the starting 
gravitational theory is Einstein gravity. Nevertheless, 
the quite natural extension is to consider the higher derivative 
gravitational theory and to study BHs and their entropy in 
such theory. Indeed, higher derivative gravity in $d$ dimensions 
naturally appears as string effective action in the sigma-model 
approach to string theory. Moreover, terms quadratic on the 
curvatures in such stringy effective action\cite{12}
very often form the Gauss-Bonnet combination which is topological 
invariant in four dimensions. The investigation of BHs in such theory 
leads to very important information. For example, it turns out that 
BH entropy in higher derivative gravity is not always 
proportional to the area of horizon \cite{22,NOOr2a}. There are 
indications \cite{NOOr2a} that higher derivative gravity may be 
the initial step in the exploring of AdS/non-CFT correspondence.

Among the different black holes the special interest from 
the holographic point of view is attracted by Schwarzschild-Anti 
de Sitter (SAdS) and Schwarzschild-de Sitter (SdS) black holes. 
Indeed, some of SAdS BHs lead to the very interesting phenomenon 
called Hawking-Page phase transition \cite{HP}.
This Hawking-Page phase transition plays an important role in AdS/CFT 
correspondence where it was interpreted by Witten\cite{AdS} as the 
confinement-deconfinement transition in dual gauge theory.
Hence, SAdS BHs are the natural tool to describe thermodynamics of
dual SCFT what gives the important support of AdS/CFT correspondence.
Moreover, the study of entropy for such SAdS BHs is not only 
fundamental by itself but it provides the important information 
about the entropy of the early Universe (again, via dual description).

 From another point, SdS BHs seem to be very useful for the
recently proposed dS/CFT correspondence which is not yet well 
understood. Moreover, as there are indications that our modern 
Universe has small positive cosmological 
constant, SdS BHs which limiting cases are de Sitter and Nariai space 
may find various cosmological applications. Furthermore, 
there are many similarities between 
SAdS and SdS BHs which probably reflect the common origin of 
AdS(dS)/CFT correspondence. 

In the present paper we consider SAdS and SdS BH thermodynamics for 
Einstein-Gauss-Bonnet gravity (with electromagnetic field).
The motivation to consider such theory is two-fold.
 From the one side, GB combination naturally appears as 
next-to-leading term in heterotic string effective action. From 
another side, this is example of the theory with higher derivative 
terms where, nevertheless, the field equations are of second order 
like in pure Einstein gravity. Hence, it provides the workable 
and analytically solvable theory which still remains higher 
derivative one and keeps many features of its more general cousins. 

In the next section the field equations for $d$-dimensional 
higher derivative gravity are written for class of metrics typical
for SAdS space. It is demonstrated that when higher derivative 
terms form GB combination the field equations become of second 
order. As the matter, the electromagnetic field is introduced.
The analytical solution in the form of SAdS BH is presented for such 
Einstein-GB-Maxwell theory. The thermodynamics of such SAdS BH is 
carefully described: the Hawking temperature, multiple horizon radius,
regularized action and free energy are found. In a sense, 
this section may be considered as higher derivative generalization 
of corresponding calculation for charged AdS BH in 
ref.\cite{mirjam} (in our case there is no dilaton).
 However, unlike to \cite{mirjam} where both canonical and 
grand-canonical ensembles were discussed we 
study the phases of charged AdS BH in only canonical ensemble.
The reason is caused by very complicated structure of the theory under 
investigation.
The Hawking-Page phase transitions are described, their dependence on
the coefficient of GB term and on the electric charge is demonstrated.
 It is 
remarkable that due to presence of above two parameters, 
the critical line (not critical point) appears in Hawking-Page phase diagram.
In section three we show the principal possibility to account this 
formulation  for higher derivative electromagnetic field as 
well. Unfortunately, the explicit calculations are getting extremely 
complicated. 

In section four we begin to consider dS bulk spaces. Using proposed 
dS/CFT correspondence, in direct analogy with AdS/CFT it is shown 
that brane gravity maybe localized even when bulk is de Sitter space.
Section five is devoted to the review of de Sitter space as well
as SdS BH thermodynamics in usual Einstein gravity. The derivation of 
thermodynamical energy (conserved black hole mass) and entropy for SdS BH is 
presented. In section six the properties of charged SdS BH for 
Einstein-GB-Maxwell theory are analyzed. In particular, the Hawking 
temperature, multiple horizon radius are found and the expression for 
conserved BH mass is conjectured. Section seven is devoted to quite 
complicated technical problem, i.e. the derivation of surface 
counterterms for higher derivative gravity on SAdS or SdS BH background.
Using surface counterterms the conserved BH mass for general higher 
derivative gravity as well as for Einstein-GB theory is found.
In particulary, it proves the conjecture of previous section.
The interesting property of higher derivative gravity is that there 
exists critical line of theory parameters where BH mass is zero for any 
SdS or SAdS BH. The entropy of such state is also zero.
In  section eight we discuss the entropies for SdS and SAdS BHs in
higher derivative gravity as well as in Einstein-GB theory.
It is shown that for some higher derivative parameters region the entropy 
formally becomes negative. However, in the situation when SAdS BH entropy is 
negative,
for the same choice of theory parameters the SdS BH entropy is still 
positive 
and vice-versa. That is why we speculate that when say SAdS BH entropy is 
negative one enters to some new kind of BH instability and transition
to stable SdS BH occurs (and vice-versa).
The summary and outlook are given in the last section.

\section{AdS Black Holes thermodynamics and Hawking-Page 
phase transitions \label{Sec2}}

We will start from the
 general action of $d$ dimensional $R^2$-gravity with 
cosmological constant and matter. The action is given by:
\be
\label{vi}
S=\int d^{d+1} x \sqrt{-g}\left\{a R^2 + b R_{\mu\nu} R^{\mu\nu}
+ c  R_{\mu\nu\xi\sigma} R^{\mu\nu\xi\sigma}
+ {1 \over \kappa^2} R - \Lambda 
\right\}+ S_{\rm matter}\ .
\ee
Here $S_{\rm matter}$ is the action for the matter fields. 
Note that in string theory the coefficients of above action are derived.
Of course, they will depend on the starting string theory and the 
compactification used. Usually, the Einstein term is the leading one 
while higher derivative terms are next-to-leading ones in low-energy 
stringy effective action. Nevertheless, even in such situation the higher
derivative terms may change the structure of Hawking-Page phase transitions
\cite{noplb}.

By the variation over the metric tensor $g_{\mu\nu}$, we obtain 
the following equation
\bea
\label{R1}
0&=&{1 \over 2}g^{\mu\nu}\left(a R^2 + b R_{\mu\nu} R^{\mu\nu}
+ c R_{\mu\nu\xi\sigma} R^{\mu\nu\xi\sigma}
+ {1 \over \kappa^2} R - \Lambda \right) \nn
&& + a\left(-2RR^{\mu\nu} + \nabla^\mu\nabla^\nu R 
+\nabla^\nu\nabla^\mu R 
 - 2g^{\mu\nu} \nabla_\rho\nabla^\rho R \right) \nn
&& + b\left( -2 R^\mu_{\ \rho} R^{\nu\rho} 
+ \nabla_\rho\nabla^\mu R^{\rho\nu}
+ \nabla_\rho\nabla^\nu R^{\rho\mu} - \Box R^{\mu\nu} 
 - g^{\mu\nu}\nabla^\rho\nabla^\sigma R_{\rho\sigma} \right) \nn
&& + c\left(-2R^{\mu\rho\sigma\tau}R^\nu_{\ \rho\sigma\tau}
 - 2 \nabla_\rho\nabla_\sigma R^{\mu\rho\nu\sigma}
 - 2 \nabla_\rho\nabla_\sigma R^{\nu\rho\mu\sigma}\right) \nn
&& - { 1 \over \kappa^2}R^{\mu\nu} + T^{\mu\nu} \ .
\eea
Here $T^{\mu\nu}$ is the energy-momentum tensor of 
the matter fields:
\be
\label{GBi}
T^{\mu\nu}={1 \over \sqrt{-g}}
{\delta S_{\rm matter} \over \delta g_{\mu\nu}}\ .
\ee
Using the following identities:
\bea
\label{GBii}
\nabla_\rho \nabla_\sigma R^{\mu\rho\nu\sigma} &=&
\Box R^{\mu\nu} - {1 \over 2}\nabla^\mu \nabla^\nu R 
+ R^{\mu\rho\nu\sigma} R_{\rho\sigma} 
 - R^\mu_{\ \rho} R^{\nu\rho} \ ,\nn
\nabla_\rho \nabla^\mu R^{\rho\nu} 
+ \nabla_\rho \nabla^\nu R^{\rho\mu} 
&=& {1 \over 2} \left(\nabla^\mu \nabla^\nu R 
+ \nabla^\nu \nabla^\mu R\right)
 - 2 R^{\mu\rho\nu\sigma} R_{\rho\sigma} 
+ 2 R^\mu_{\ \rho} R^{\nu\rho} \ ,\nn
\nabla_\rho \nabla_\sigma R^{\rho\sigma} &=& {1 \over 2} R \ ,
\eea
one rewrites Eq.(\ref{R1}) as:
\bea
\label{R2}
0&=&{1 \over 2}g^{\mu\nu}\left(a R^2 + b R_{\mu\nu} R^{\mu\nu}
+ c R_{\mu\nu\xi\sigma} R^{\mu\nu\xi\sigma}
+ {1 \over \kappa^2} R - \Lambda \right) \nn
&& + a\left(-2RR^{\mu\nu} + \nabla^\mu\nabla^\nu R 
+\nabla^\nu\nabla^\mu R 
 - 2g^{\mu\nu} \nabla_\rho\nabla^\rho R \right) \nn
&& + b\left\{ {1 \over 2} \left(\nabla^\mu \nabla^\nu R 
+ \nabla^\nu \nabla^\mu R\right)
 - 2 R^{\mu\rho\nu\sigma} R_{\rho\sigma} 
 - \Box R^{\mu\nu} - {1 \over 2} g^{\mu\nu}\Box R \right\} \nn
&& + c\left(-2R^{\mu\rho\sigma\tau}R^\nu_{\ \rho\sigma\tau}
 - 4 \Box R^{\mu\nu} + \nabla^\mu \nabla^\nu R 
+ \nabla^\nu \nabla^\mu R \right. \nn
&& \left. - 4 R^{\mu\rho\nu\sigma} R_{\rho\sigma}
+ 4 R^\mu_{\ \rho} R^{\nu\rho} \right) \nn
&& - { 1 \over \kappa^2}R^{\mu\nu} - T_{\rm matter}^{\mu\nu} \ .
\eea
Our primary interest is in the study of the case when 
 the $R^2$-part of the action (\ref{vi}) forms 
the Gauss-Bonnet combination:
\be
\label{GBiii}
a=c\ ,\quad b=-4c\ ,
\ee
Then
\bea
\label{R3}
0&=&{1 \over 2}g^{\mu\nu}\left\{c\left( R^2 - 4 R_{\mu\nu} R^{\mu\nu}
+  R_{\mu\nu\xi\sigma} R^{\mu\nu\xi\sigma}\right)
+ {1 \over \kappa^2} R - \Lambda \right\} \nn
&& + c\left(-2RR^{\mu\nu} + 4 R^\mu_{\ \rho} R^{\nu\rho} 
+ 4 R^{\mu\rho\nu\sigma} R_{\rho\sigma} 
 - 2 R^{\mu\rho\sigma\tau}R^\nu_{\ \rho\sigma\tau} \right) \nn
&& - { 1 \over \kappa^2}R^{\mu\nu} + T^{\mu\nu} \ .
\eea
 Eq. (\ref{R3}) does not contain the 
derivatives of the curvatures therefore the terms with the 
derivatives higher than two do not appear. Therefore the 
theory with Gauss-Bonnet combination is very special case of higher 
derivative gravity. 

We now assume the spacetime metric has the following form:
\be
\label{GBiv}
ds^2 = - \e^{2\nu (r)} dt^2 + \e^{2\lambda (r)} dr^2 
+ r^2 \sum_{i,j=1}^{d-1} \tilde g_{ij} dx^i dx^j\ .
\ee
Here $\tilde g_{ij}$ is the metric of the Einstein manifold, 
which is defined by $\tilde R_{ij}=k g_{ij}$. Here 
$\tilde R_{ij}$ is the Ricci curvature given by $\tilde g_{ij}$ 
and $k$ is a constant. For example, we have $k=d-2$ for 
$d-1$-dimensional unit sphere, $k=-(d-2)$ for $d-1$-dimensional 
unit hyperboloid, and $k=0$ for flat surface. 

For the metric (\ref{GBiv}), the non-vanishing   
curvatures are:
\bea
\label{GBv}
R_{rtrt}&=&\e^{2\nu}\left\{\nu'' + \left(\nu' - \lambda'\right)
\nu'\right\} \ ,\nn
R_{titj}&=& r\nu'\e^{2(\nu - \lambda)} \tilde g_{ij} \ ,\nn
R_{rirj}&=& r\lambda' \tilde g_{ij} \ ,\nn
R_{ijkl}&=& \left({k \over d-2} - \e^{-2\lambda}\right) r^2
\left(\tilde g_{ik} \tilde g_{jl} - \tilde g_{il} \tilde g_{jk} 
\right)\ ,\nn
R_{tt}&=& \e^{2\left(\nu - \lambda\right)} \left\{
\nu'' + \left(\nu' - \lambda'\right)\nu' 
+ {(d-1) \nu' \over r}\right\} \ ,\nn
R_{rr}&=& - \left\{ \nu'' + \left(\nu' - \lambda'\right)\nu' 
\right\} + {(d-1) \lambda' \over r} \ ,\nn
R_{ij}&=& \left[ k + \left\{ -d + 2 - r \left(\nu' - 
\lambda' \right)\right\}\e^{-2\lambda}\right]
\tilde g_{ij}\ , \nn
R&=& \e^{-2\lambda}\left[ - 2\nu'' - 2\left(\nu'
 - \lambda'\right)\nu' - {2(d-1)\left(\nu'
 - \lambda'\right) \over r} \right. \nn
&& \left. + {(d-1)k\e^{2\lambda} 
 - (d-1)(d-2) \over r^2} \right] \ .
\eea

In (\ref{GBv}), we denote the derivative with respect to 
$r$ by $'$ and we use the following conventions of curvatures: 
\bea
\label{curv}
R&=&g^{\mu\nu}R_{\mu\nu}\ , \nn
R_{\mu\nu}&=& {R^\lambda}_{\mu\lambda\nu}\ , \nn
{R^\lambda}_{\mu\rho\sigma}&=& -\Gamma^\lambda_{\mu\rho,\nu}
+ \Gamma^\lambda_{\mu\nu,\rho}
 - \Gamma^\eta_{\mu\rho}\Gamma^\lambda_{\nu\eta}
+ \Gamma^\eta_{\mu\nu}\Gamma^\lambda_{\rho\eta}\ , \nn
\Gamma^\eta_{\mu\lambda}&=&{1 \over 2}g^{\eta\nu}\left(
g_{\mu\nu,\lambda} + g_{\lambda\nu,\mu} - g_{\mu\lambda,\nu} 
\right)\ .
\eea
Then $(\mu,\nu)=(t,t)$, $(r,r)$ and $(i,j)$ components of 
Eq.(\ref{R3}) are: 
\bea
\label{GBvi}
0&=& - {\e^{-2\nu} \over 2}\left[ -c \e^{-2\lambda}\K 
\left\{ - {4(d-1)(d-3)\lambda' \over r^3} \right.\right.\nn
&& \left. - {(d-1)(d-3)(d-4)\K \e^{2\lambda} \over 
(d-2)r^4}\right\} \\
&& \left. + {\e^{-2\lambda} \over \kappa^2}\left\{
{2(d-1)\lambda' \over r} + {(d-1)\K \e^{2\lambda} \over r^2} 
\right\} - \Lambda\right] + T^{tt} \ , \nn
\label{GBvii}
0&=& {\e^{-2\lambda} \over 2}\left[ -c \e^{-2\lambda}\K 
\left\{ {4(d-1)(d-3)\nu' \over r^3} \right.\right. \nn
&& \left. - {(d-1)(d-3)(d-4)\K \e^{2\lambda} \over 
(d-2)r^4}\right\} \\
&& \left. + {\e^{-2\lambda} \over \kappa^2}\left\{
 - {2(d-1)\nu' \over r} + {(d-1)\K \e^{2\lambda} \over r^2} 
\right\} - \Lambda\right] + T^{rr} \ , \nn
\label{GBviii}
0&=& {1 \over 2r^2}\tilde g^{ij} \left[ -c\e^{-2\lambda} 
\left\{ \K \left( {4(d-3) \over r^2}\left(\nu'' 
+ \left(\nu' - \lambda'\right)\nu'\right) 
\right.\right.\right. \nn
&& + {4(d-2)(d -3)\left(\nu' - \lambda'\right) \over r^3 } \nn
&& \left. + {\left(d-3\right)\left(d-4\right)\left(d-5 \right) 
\over (d-2) r^4}\K \e^{2\lambda} \right) \nn
&& \left. + {8(d-3)(d-2) \over r^2}\nu'\lambda'\right\} \nn
&& + {\e^{-2\lambda} \over \kappa^2}\left\{ - 2\nu'' 
 -2 \left(\nu' - \lambda'\right)\nu' 
 - {2(d-1)\left(\nu' - \lambda'\right) \over r} \right. \nn
&& \left.\left. + {(d+2) \over r^2}\K \e^{2\lambda}\right\}
 - \Lambda\right] + T^{ij}\ .
\eea
Especially for $k=2$, we have
\bea
\label{GBvid2}
0&=& - {\e^{-2\nu} \over 2}\left[ -c \e^{-2\lambda}
(d-1)(d-2)(d-3)\left(1-\e^{-2\lambda}\right)
\left\{ - {4\lambda' \over r^3} \right.\right.\nn
&& \left. - {(d-4)\left(1-\e^{-2\lambda}\right) \e^{2\lambda} 
\over r^4}\right\} \\
&& \left. + {(d-1)\e^{-2\lambda} \over \kappa^2}\left\{
{2\lambda' \over r} + {(d-2)\left(1-\e^{-2\lambda}\right) 
\e^{2\lambda} \over r^2} \right\} - \Lambda\right] + T^{tt} \ , \nn
\label{GBviid2}
0&=& {\e^{-2\nu} \over 2}\left[ -c \e^{-2\lambda}
(d-1)(d-2)(d-3)\left(1-\e^{-2\lambda}\right)
\left\{ {4\nu \over r^3} \right.\right.\nn
&& \left. - {(d-4)\left(1-\e^{-2\lambda}\right) \e^{2\lambda} 
\over r^4}\right\} \\
&& \left. + {(d-1)\e^{-2\lambda} \over \kappa^2}
\left\{-{2\nu' \over r} + {(d-2)\left(1-\e^{-2\lambda}\right) 
\e^{2\lambda} \over r^2} \right\} - \Lambda\right] + T^{rr} \ , \nn
\label{GBviiid2}
0&=& {1 \over 2r^2}\tilde g^{ij} \left[ -c\e^{-2\lambda} 
\left\{ (d-2)\left(1-\e^{-2\lambda}\right)
 \left( {4(d-3) \over r^2}\left(\nu'' 
+ \left(\nu' - \lambda'\right)\nu'\right) 
\right.\right.\right. \nn
&& + {4(d-2)(d -3)\left(\nu' - \lambda'\right) \over r^3 } 
\left. + {\left(d-3\right)\left(d-4\right)\left(d-5\right) 
\over r^4}\left(1-\e^{-2\lambda}\right) \e^{2\lambda} \right) \nn
&& \left. + {8(d-3)(d-2) \over r^2}\nu'\lambda'\right\} \nn
&& + {\e^{-2\lambda} \over \kappa^2}\left\{ - 2\nu'' 
 -2 \left(\nu' - \lambda'\right)\nu' 
 - {2(d-1)\left(\nu' - \lambda'\right) \over r} \right. \nn
&& \left.\left. + {(d+2)(d-2) \over r^2}
\left(1-\e^{-2\lambda}\right) \e^{2\lambda}\right\}
 - \Lambda\right] + T^{ij}\ .
\eea
Combining (\ref{GBvi}) and (\ref{GBvii}), one gets
\bea
\label{GBix}
0&=& {\e^{-2\lambda} \over 2}\left[ {4(d-1)(d-3)c \over r^3}\K 
 + {2(d-1) \over \kappa^2 r}\right]\left(\nu' + \lambda'\right) \nn
&& + \e^{2\nu}T^{tt} + \e^{2\lambda}T^{rr}\ .
\eea
Therefore if
\be
\label{GBx}
0=\e^{2\nu}T^{tt} + \e^{2\lambda}T^{rr}\ ,
\ee
then $\nu' + \lambda' =0$, that is, $\nu + \lambda =$constant. 
The constant can be, as usually, absorbed into the redefinition of 
the time variable $t$ and 
\be
\label{GBxi}
\nu=-\lambda\ .
\ee
Then Eq.(\ref{GBvi}) can be rewritten as 
\bea
\label{GBxii}
0&=&{d \over dr}\left[(d-1)(d-2)c\left\{ - 2k r^{d-4}\e^{-2\lambda} 
+ (d-2)r^{d-4}\e^{-4\lambda}\right\} \right.\nn
&& \left. - {(d-1) \over \kappa^2} r^{d-2}\e^{-2\lambda}\right] 
 + {(d-1)(d-3)(d-4) \over d-2}ck^2 r^{d-5} \nn
&& + {(d-1)k \over \kappa^2}kr^{d-3} - \Lambda r^{d-1} 
+ 2r^{d-1} \e^{2\lambda} T^{rr}\ .
\eea
Let us consider the electromagnetic field, as a matter field:
\bea
\label{GBxiii}
S_{\rm matter}&=& -{1 \over 4g^2}\int d^{d+1}x \sqrt{-g} 
g^{\mu\nu}g^{\rho\sigma} F_{\mu\rho} F_{\nu\sigma}\ , \nn
F_{\mu\nu}&=& \partial_\mu A_\nu - \partial_\nu A_\mu \ .
\eea
Here $A_\mu$ is a vector potential (gauge field) and $g$ is a 
gauge coupling. Then the energy-momentum tensor is given by
\be
\label{GBxiv}
T^{\mu\nu}=-{1 \over 4g^2}\left({1 \over 2}g^{\mu\nu}g^{\rho\sigma}
g^{\xi\eta} F_{\rho\xi} F_{\sigma\eta} 
 - 2 g_{\rho\sigma}F^{\mu\rho}F^{\nu\sigma}\right)\ ,
\ee
and by the variation over $A_\mu$, one arrives at the following equation:
\be
\label{GBxv}
\partial_\nu\left(\sqrt{-g}F^{\nu\mu}\right)=0\ .
\ee
Assume $F_{tr}=-F_{rt}$ only depends on $r$ and other 
components in the field strength $F_{\mu\nu}$ vanish. Then 
 using (\ref{GBxv}), we find 
\be
\label{GBxvi}
0=\partial_r\left(\e^{-\nu-\lambda}r^{d-1} F_{tr}\right)\ ,
\ee
whose solution is given by
\be
\label{GBxvii}
F_{tr}=\e^{\nu+\lambda}r^{1-d}Q\ .
\ee
Here $Q$ is a constant corresponding to the charge. 
Then the energy-momentum tensor (\ref{GBxiv}) is
\be
\label{GBxviii}
T^{tt}= {\e^{2\nu} r^{2-2d} Q^2 \over 4g^2}\ ,\quad 
T^{rr}= - {\e^{2\lambda} r^{2-2d} Q^2 \over 4g^2}\ ,\quad 
T^{ij}= { r^{-2d} Q^2 \over 4g^2}\tilde g^{ij}\ .
\ee
 Eq.(\ref{GBx}) is satisfied and  $\nu=-\lambda$ in 
(\ref{GBxi}). From Eq.(\ref{GBxii}), we find
\bea
\label{GBxix}
0&=&(d-1)(d-2)c\left\{ - 2k r^{d-4}\e^{-2\lambda} 
+ (d-2)r^{d-4}\e^{-4\lambda}\right\} 
 - {(d-1) \over \kappa^2} r^{d-2}\e^{-2\lambda} \nn
&& + {(d-1)(d-3) \over d-2}ck^2 r^{d-4} 
+ {(d-1)k \over (d-2)\kappa^2}r^{d-2} \nn
&& - {\Lambda \over d}r^d 
+ {(d-2)Q^2 \over 2g^2}r^{2-d} + C\ .
\eea
Here $C$ is an integration constant related with mass.
Then one can solve (\ref{GBxix}):
\bea
\label{GBxx}
\e^{2\nu}&=&\e^{-2\lambda} \nn
&=& {1 \over 2c}\left[{2ck \over d-2} + {r^2 \over 
\kappa^2 (d-2)(d-3)} 
\right.\nn
&& \pm \left\{ {r^4 \over \kappa^4 (d-2)^2(d-3)^2} 
+ {4c \Lambda r^4 \over d(d-1)(d-2)(d-3)} \right. \nn
&& \left.\left. - {2cQ^2 r^{6-2d} \over g^2 (d-1)(d-3)} 
 - {4cC r^{4-d} \over (d-1)(d-2)(d-3)} 
 \right\}^{1 \over 2}\right]\ .
\eea
When $Q^2=0$, the above solution reproduces the result in 
\cite{Cai},\footnote{
There is some difference in the notations  here and in 
\cite{Cai}. For example $k$ in \cite{Cai} is ${k \over d-1}$ 
in this paper and $d$ in \cite{Cai} is $d+1$.}
Then from (\ref{GBxix}), we have the horizon, where $\e^{2\nu}=0$, 
at $r=r_H$ 
\bea
\label{GBxxi}
0 &=& {(d-1)(d-3) \over d-2}ck^2 r_H^{d-4} 
+ {(d-1)k \over (d-2)\kappa^2}r_H^{d-2} \nn
&& - {\Lambda \over d}r_H^d 
+ {(d-2)Q^2 \over 2g^2}r_H^{2-d} + C\ ,
\eea
and the Hawking temperature $T_H$ is given by
\bea
\label{GBxxii}
4\pi T_H &=& \left.\left(\e^{2\nu}\right)'\right|_{r=r_H} \nn
&=& {1 \over 2}\left({2ck \over d-2} + {r_H^2 \over \kappa^2 (d-2)(d-3)}
\right)^{-1}\left[ {4kr_H \over \kappa^2 (d-2)^2(d-3)} \right. \nn
&& - {8\Lambda r_H^3 \over d(d-1)(d-2)(d-3)} 
 - {2 Q^2 r_H^{5-2d} \over (d-1)g^2} \nn
&& \left.  
 - {2(d-4) C r_H^{3-d} \over (d-1)(d-2)(d-3)} \right]\ .
\eea

We now concentrate on $d=4$ case and consider the 
thermodynamics. When $d=4$, Eq.(\ref{GBxx}) has the 
following form:
\bea
\label{GBxxiii}
\e^{2\nu}&=&\e^{-2\lambda} \nn
&=& {1 \over 2c}\left[ck + {r^2 \over 2\kappa^2 } 
\pm \left\{ {r^4 \over 4\kappa^4} 
+ {c \Lambda r^4 \over 6}  - {2cQ^2 \over 3g^2 r^2 } 
 - {2c\tilde C \over 3} \right\}^{1 \over 2}\right]\ .
\eea
Here
\be
\label{GBai}
\tilde C = C + {3 \over 2}ck^2\ .
\ee
Then the asymptotic behavior when $r$ is large is given by
\bea
\label{GBxxiv}
\e^{2\nu}&=&\e^{-2\lambda} \nn
&=& {1 \over 2c}\left[{r^2 \over 2\kappa^2 }\left(1\pm 
\sqrt{1 + {2c\Lambda \kappa^4 \over 3}}\right) +ck \right. \nn
&& \left. \mp {2\kappa^2 c \tilde C \over 3r^2
\sqrt{1 + {2c\Lambda \kappa^4 \over 3}}}
\mp {2cQ^2 \over 3g^2 r^4 
\sqrt{1 + {2c\Lambda \kappa^4 \over 3}}}
+ {\cal O}\left(r^{-4}\right) \right]\ .
\eea
One can compare the above behavior (\ref{GBxxiv}) with  
that of the usual Reissner-Nordstr\o m-AdS case
\be
\label{GBxxv}
\e^{2\nu_{\rm SAdS}}=\e^{-2\lambda_{\rm SAdS}} 
= {r^2 \over l^2} + {k \over 2} - {\mu \over r^2} 
+ {q^2 \over r^4}\ ,
\ee
We identify
\bea
\label{GBxxvi}
&&{1 \over l^2}={1 \over 4c\kappa^2 }\left(1\pm 
\sqrt{1 + {2c\Lambda \kappa^4 \over 3}}\right) \ ,\quad 
\mu=\pm {\kappa^2 \tilde C \over 3
\sqrt{1 + {2c\Lambda \kappa^4 \over 3}}}\ ,\nn
&& q^2=\mp {Q^2 \over 3g^2\sqrt{1 + {2c\Lambda \kappa^4 \over 3}}}\ ,
\eea
that is, 
\bea
\label{GBxxvii}
&& \Lambda = - {12 \over \kappa^2 l^2} + {24c \over l^4}\ ,\quad 
\tilde C=\mu\left({12c \over l^2} - {3 \over \kappa^2}\right)\ ,\nn
&& {Q^2 \over g^2}=3q^2 \left(1 - 
{4c\kappa^2 \over l^2}\right)\ . 
\eea
Eq.(\ref{GBxxvi}) tells that the plus (minus) sign in $\pm$ of 
(\ref{GBxxiii}) corresponds to the case of ${c\kappa^2 \over l^2}>{1 \over 
4}$ $\left({c\kappa^2 \over l^2}<{1 \over 4}\right)$. 
 Eq.(\ref{GBxxvii}) tells $q^2$ is negative 
when ${c\kappa^2 \over l^2}>{1 \over 4}$. Then 
the charge itself appears to be imaginary for the observer far from 
the black hole. 

By using $l$ and $\mu$ in (\ref{GBxxvii}) instead of $\Lambda$ and 
$C$, one rewrites (\ref{GBxxiii}), (\ref{GBxxi}) and 
(\ref{GBxxii}) (for $d=4$) in the following form (we prefer $Q^2$ 
rather than $q^2$ since $q^2$ becomes negative 
when ${c\kappa^2 \over l^2}>{1 \over 4}$):
\bea
\label{GBxxviii}
\e^{2\nu}&=&\e^{-2\lambda} \nn
&=& {1 \over 2c}\left\{ck + {r^2 \over 2\kappa^2 } \right.\nn
&& \left. \pm \sqrt{ {r^4 \over 4\kappa^4}
\left({4c\kappa^2 \over l^2} -1
\right)^2 - {2c\mu \over \kappa^2}\left({4c\kappa^2 \over l^2} -1
\right) - {2cQ^2 \over 3g^2r^2 }} \right\}\ ,\\
\label{GBxxix}
0 &=& r_H^6 - {k \over 2\kappa^2
\left({2c \over l^4} - {1 \over \kappa^2 
l^2}\right)}r_H^4 - {\mu \left( 
{4c \over l^2} - {1 \over \kappa^2}\right) \over 
{2c \over l^4} - {1 \over \kappa^2 l^2}}r_H^2 \nn
&& - {Q^2 \over 3g^2 \left({2c \over l^4} - {1 \over \kappa^2 
l^2}\right)}\ ,\\
\label{GBxxx}
4\pi T_H &=& \left.\left(\e^{2\nu}\right)'\right|_{r=r_H} \nn
&=& {1 \over 2}\left(ck + {r_H^2 \over 2\kappa^2 }
\right)^{-1}\left[ {kr_H \over \kappa^2 } 
 - {8c r_H^3 \over l^4}
 + {4 r_H^3 \over \kappa^2 l^2} - {2 Q^2 \over 3 g^2 r_H^3} 
\right]\ .
\eea
Eq.(\ref{GBxxix}) has two positive solutions for $r_H^2$, 
that is, the black hole has two horizons if 
\bea
\label{GBxxxi}
{Q^2 \over g^2}&<& {Q_c^2 \over g^2} \nn
&\equiv& -{1 \over 9}\left({2c \over l^4} - {1 \over \kappa^2 
l^2}\right)^{-2}\left[ {k^2 \over 4} 
+ 3 \mu \left({4c \over l^2}  - {1 \over \kappa^2 }\right)
\left({2c \over l^4} - {1 \over \kappa^2 
l^2}\right)\right]^{3 \over 2} \nn
&& - {k\mu\left({4c \over l^2} 
 - {1 \over \kappa^2 }\right) \over 3\left({2c \over l^4} 
 - {1 \over \kappa^2 l^2}\right)}
+ {k^3 \over 72 \left({2c \over l^4} 
 - {1 \over \kappa^2 l^2}\right)^2}\ .
\eea
The case of $Q^2=Q_c^2$ corresponds to the extremal case. 
The explicit solutions of (\ref{GBxxix}) are given by
\bea
\label{GBxxxib} 
r_H^2&=& {\tilde a \over 3} + \alpha_+ + \alpha_- \ ,
\quad {\tilde a \over 3} + \alpha_+ \zeta + \alpha_- \zeta^2 \ ,
\quad {\tilde a \over 3} + \alpha_+ \zeta^2 + \alpha_- \zeta 
\ ,\nn
\zeta&=&\e^{2\pi i \over 3} \ ,\nn
\alpha_\pm^3&=& {1 \over 4}\left\{ - \left(
{\tilde a^3 \over 27} + \tilde c \right) \pm 
\sqrt{\left({\tilde a^3 \over 27} + \tilde c\right)^2 
 - {4 \over 27}\left({\tilde a^2 \over 3} - \tilde b\right)^3}
\right\}\ ,\nn
\tilde a &=& - {k \over 2\left({2c \over l^4} - {1 \over \kappa^2 
l^2}\right)}\ ,\nn
\tilde b &=& - {\mu \left( 
{4c \over l^2} - {1 \over \kappa^2}\right) \over 
{2c \over l^4} - {1 \over \kappa^2 l^2}}\ ,\nn
\tilde c &=& - {Q^2 \over 3g^2 \left({2c \over l^4}
 - {1 \over \kappa^2 l^2}\right)}\ .
\eea
After Wick-rotating the time variable by $t \to i\tau$, the
free energy $F$ can be obtained from the action $S$ in 
(\ref{vi}) with $a=c$ and $b=-4$ and $S_{\rm matter}$ in 
(\ref{GBxiii}), where the classical solution is substituted:
\be
\label{GBxxxii}
F=-TS\ .
\ee
Multiplying $g^{\mu\nu}$ to (\ref{R3}) with (\ref{GBxiv}), we 
obtain
\be
\label{GBxxxiii}
0=c\left( R^2 - 4 R_{\mu\nu} R^{\mu\nu}
+  R_{\mu\nu\xi\sigma} R^{\mu\nu\xi\sigma}\right)
+ {3 \over \kappa^2}R - 5\Lambda 
 -{1 \over 4g^2} g^{\mu\nu}g^{\rho\sigma} 
F_{\mu\rho} F_{\nu\sigma}\ .
\ee
Then the action can be rewritten in the following form 
\be
\label{GBxxxiv}
S=\int d^5x \sqrt{g}\left( - {2 \over \kappa^2}R + 4\Lambda \right)
={V_3 \over T_H}\int_{r_H}^\infty dr r^3
\left( - {2 \over \kappa^2}R + 4\Lambda \right)\ .
\ee
Here $V_{3}$ is the volume of 3d sphere and  $\tau$ 
has a period  ${1\over T_H}$. The expression for $S$ contains 
the divergence coming from large $r$. In order to subtract the 
divergence, we regularize $S$ (\ref{GBxxxiv}) by cutting off the 
integral at a large radius $r_{\rm max}$ and subtracting the 
solution with $\mu = Q = 0$: 
\bea
\label{GBxxxv}
S_{\rm reg}&=& 
-{V_{3} \over T}\left\{ \int ^{r_{\rm max}}_{r_H} dr r^3
\left( - {2 \over \kappa^2}R + 4\Lambda \right)\right. \\
&& \left. -\e^{-\lambda(r=r_{\rm max})+\lambda(r=r_{\rm max};
\mu = Q = 0) }
\int ^{r_{\rm max}}_0 dr r^3
\left.\left(  - {2 \over \kappa^2}R + 4\Lambda 
\right) \right|_{\mu = Q = 0} 
\right\}\ .\nonumber
\eea
The factor $\e^{-\lambda(r=r_{\rm max})
+ \lambda(r=r_{\rm max};\mu = Q = 0)}$ 
is chosen so that the proper length of the circle which 
corresponds to the period ${1 \over T_H}$ in the Euclidean 
time at $r=r_{\rm max}$ coincides with each other in the 
two solutions. 
We should note that the scalar curvature $R$ in (\ref{GBv}) with 
$\nu=-\lambda$ has the following form 
\be
\label{GBxxxvi}
R=-{1 \over r^3}\left(r^3\e^{-2\lambda}\right)' 
+ {3k \over r^2}
\ee
and $\e^{-2\lambda}=0$ and $\left(\e^{-2\lambda}\right)'=4\pi 
T_H$ at the horizon $r=r_H$. Then in the limit of 
$r_{\rm max}\rightarrow \infty$, we find the following 
expression of the free energy
\be
\label{GBxxxvii}
F=-V_3\left[ - {3\mu \over \kappa^2} + {12c\mu \over l^2}
 - {8\pi r_H^3 T_H - 3k r_H^2 \over \kappa^2} 
 - {24c r_H^4 \over l^4} + {12 r_H^4 \over 
\kappa^2 l^2}\right]\ .
\ee
 From the parameters $\mu$, $Q$, $r_H$, 
and $T_H$, only two parameters are independent. In fact, 
by using (\ref{GBxxix}) and (\ref{GBxxx}), one finds
\bea
\label{GBxxxviii}
\mu &=& \left({4c \over l^2} 
 - {1 \over \kappa^2}\right)^{-1}\left\{ 
 3\left({2c \over l^4} - {1 \over \kappa^2 l^2}\right)r_H^4
 - {k \over \kappa^2}r_H^2 
 + 4\pi T_H r_H\left(ck + {r_H^2 \over 2\kappa^2}\right)
 \right\} \nn
{Q^2 \over g^2}&=& -6\left({2c \over l^4} - {1 \over \kappa^2 l^2}
\right) r_H^6 + {3k \over 2\kappa^2}r_H^4 - 12\pi T_H r_H^3
\left(ck + {r_H^2 \over 2\kappa^2}\right)\ .
\eea
Using (\ref{GBxxxviii}), we can express the free energy 
$F$  (\ref{GBxxxvii}) 
in terms of $r_H$ and $T_H$:
\be
\label{GBxxxix}
F=-V_3\left\{ -3 \left({2c \over l^4} - {1 \over \kappa^2 l^2}
\right)r_H^4 - 2\pi T_H r_H \left({r_H^2 \over \kappa^2} 
 - 6ck \right)\right\}\ .
\ee
Then there is a critical line, where $F=0$, when 
\be
\label{GBxxxx}
T_H=T_c \equiv {3 \over 2\pi} \left({r_H^2 \over \kappa^2} 
 - 6ck \right)^{-1} \left({1 \over \kappa^2 l^2}
 - {2c \over l^4} \right)r_H^3 \ .
\ee
Then if $T_H>T_c$ and $r_H^2 > 6 ck \kappa$ 
(or $T_H<T_c$ and $r_H^2 < 6 ck \kappa$), the pure AdS spacetime 
is more stable than the black hole spacetime and 
if $T_H<T_c$ and $r_H^2 > 6 ck \kappa$ 
(or $T_H>T_c$ and $r_H^2 < 6 ck \kappa$), vice versa. 

We now investigate the phase structure in more details. 
For this purpose, we will define a parameter $\epsilon$ as
\be
\label{GBivxiv}
\epsilon\equiv {c\kappa^2 \over l^2}\ ,
\ee
and rescale $r_H$ and $T_H$ by
\be
\label{GBivxv}
r_H\rightarrow lr_H\ ,\quad T_H\rightarrow {T_H \over l}\ .
\ee
The critical temperature $T_c$ in (\ref{GBxxxx}) 
can be rewritten as
\be
\label{GBivxvi}
T_c={3\left(1 - 2\epsilon\right)
r_H^3 \over 2\pi \left(r_H^2 - 6k\epsilon\right)}\ .
\ee
We now assume $k=d-2=2>0$. 
When $k=2$, Eqs.(\ref{GBxxviii}), (\ref{GBxxix}), 
(\ref{GBxxx}), (\ref{GBxxxviii}), (\ref{GBxxxix}), and 
(\ref{GBivxvi}) have the following forms:
\bea
\label{GBxxviiid2}
\e^{2\nu}&=&\e^{-2\lambda} \nn
&=& {1 \over 2c}\left\{2c + {r^2 \over 2\kappa^2 } \right.\nn
&& \left. \pm \sqrt{ {r^4 \over 4\kappa^4}
\left({4c\kappa^2 \over l^2} -1
\right)^2 - {2c\mu \over \kappa^2}\left({4c\kappa^2 \over l^2} -1
\right) - {2cQ^2 \over 3g^2r^2 }} \right\}\ ,\\
\label{GBxxixd2}
0 &=& r_H^6 - {r_H^4 \over 2\kappa^2
\left({2c \over l^4} - {1 \over \kappa^2 
l^2}\right)} - {\mu \left( 
{4c \over l^2} - {1 \over \kappa^2}\right)r_H^2 \over 
{2c \over l^4} - {1 \over \kappa^2 l^2}}
 - {Q^2 \over 3g^2 \left({2c \over l^4} - {1 \over \kappa^2 
l^2}\right)}\ ,\\
\label{GBxxxd2}
4\pi T_H &=& \left.\left(\e^{2\nu}\right)'\right|_{r=r_H} \nn
&=& {1 \over 2}\left(2c + {r_H^2 \over 2\kappa^2 }
\right)^{-1}\left[ {2r_H \over \kappa^2 } 
 - {8c r_H^3 \over l^4}
 + {4 r_H^3 \over \kappa^2 l^2} - {2 Q^2 \over 3 g^2 r_H^3} 
\right]\ ,\\
\label{GBxxxviiid2}
\mu &=& \left({4c \over l^2} 
 - {1 \over \kappa^2}\right)^{-1}\left\{ 
 3\left({2c \over l^4} - {1 \over \kappa^2 l^2}\right)r_H^4
 - {2r_H^2 \over \kappa^2} 
 + 4\pi T_H r_H\left(ck + {r_H^2 \over 2\kappa^2}\right)
 \right\} \nn
{Q^2 \over g^2}&=& -6\left({2c \over l^4} - {1 \over \kappa^2 l^2}
\right) r_H^6 + {3r_H^4 \over \kappa^2} - 12\pi T_H r_H^3
\left(2c + {r_H^2 \over 2\kappa^2}\right)\ ,\\
\label{GBxxxixd2}
F&=&-V_3\left\{ -3 \left({2c \over l^4} - {1 \over \kappa^2 l^2}
\right)r_H^4 - 2\pi T_H r_H \left({r_H^2 \over \kappa^2} 
 - 12c \right)\right\}\ ,\\
\label{GBivxvid2}
T_c&=&{3\left(1 - 2\epsilon\right)
r_H^3 \over 4\pi \left(r_H^2 - 12\epsilon\right)}\ .
\eea
Then we have the following phase structure:
\begin{itemize}
\item When $\epsilon < 0$ $\left(<{1 \over 2}\right)$, 
$T_c$ in (\ref{GBivxvi}) is always positive. 
Then if $T_H>T_c$ $\left(T_H<T_c\right)$, the pure AdS 
(black hole) spacetime is more stable than black hole 
(pure AdS) spacetime. 
\item When $0<\epsilon < {1 \over 2}$, the critical 
temperature $T_c$ is positive when $r_H^2>12\epsilon$ 
and there is a critical line, where if $T_H>T_c$ 
$\left(T_H<T_c\right)$, the pure AdS (black hole) 
spacetime is more stable than the black hole (pure AdS) spacetime. 
When $r_H^2 < 12\epsilon $, $T_c$ 
is negative, then the black hole spacetime is always stable. 
\item When $\epsilon > {1 \over 2}$, 
if $r_H^2>12\epsilon$, $T_c$ is negative and 
the pure AdS spacetime is more stable than 
the black hole spacetime. If $r_H^2 < 12\epsilon $, $T_c$ is 
positive and if 
$T_H>T_c$ $\left(T_H<T_c\right)$, the black hole 
(pure AdS) spacetime is more stable than the pure AdS 
(black hole) spacetime. 
\end{itemize}

The conceptual Hawking-Page phase diagrams are given in 
Figure \ref{Fig1} for $\epsilon<{1 \over 2}$ case and in 
Figure \ref{Fig2} for $\epsilon>{1 \over 2}$ case. 

\unitlength=0.5mm

\begin{figure}
\begin{picture}(300,150)
\thicklines
\put(8,133){$T_H$}
\put(122,48){$r_H$}
\put(10,50){\vector(1,0){110}}
\put(10,50){\vector(0,1){80}}
\qbezier[400](10,50)(60,50)(119,129)
\put(20,120){pure AdS phase}
\put(60,60){AdS black hole phase}
\put(30,140){$\epsilon<0$ case}

\put(138,133){$T_H$}
\put(252,48){$r_H$}
\put(140,50){\vector(1,0){110}}
\put(140,50){\vector(0,1){80}}
\qbezier[300](181,129)(181,30)(249,129)
\put(160,60){AdS black hole phase}
\put(185,120){pure AdS}
\put(195,110){phase}
\put(165,35){$r_H^2=12\epsilon $}
\put(160,140){$0<\epsilon<{1 \over 2}$ case}
\thinlines
\put(180,50){\line(0,1){80}}
\put(180,43){\vector(0,1){7}}
\end{picture}
\caption{The phase diagrams when $\epsilon<{1 \over 2}$}
\label{Fig1}
\end{figure}

\begin{figure}
\begin{picture}(300,150)
\thicklines
\put(38,133){$T_H$}
\put(152,48){$r_H$}
\put(40,50){\vector(1,0){110}}
\put(40,50){\vector(0,1){80}}
\qbezier[200](40,50)(79,80)(79,129)
\put(75,60){pure AdS phase}
\put(48,120){AdS}
\put(46,110){black}
\put(48,100){hole}
\put(46,90){phase}
\put(75,35){$r_H^2=12\epsilon $}
\thinlines
\put(80,50){\line(0,1){80}}
\put(80,43){\vector(0,1){7}}
\end{picture}
\caption{The phase diagrams when $\epsilon>{1 \over 2}$}
\label{Fig2}
\end{figure}

We should note that the above non-trivial phase diagrams 
can be obtained since there is a charge. For the case that 
the charge vanishes, the Hawking temperature $T_H$ can be obtained 
as a function of the horizon radius $r_H$: 
\be
\label{TH1}
T_H={2\left(2\epsilon - 1\right) r_H^3 - kr_H \over 2\pi 
\left(r_H^2 + 2k\epsilon\right)}\ ,
\ee
which can be obtained from (\ref{GBxxxviii}) by putting $Q=0$ 
and by using the rescaling in (\ref{GBivxiv}) 
and (\ref{GBivxv}). 
Then although there is a critical point corresponding to Hawking-Page 
phase transition \cite{HP}, 
there does not appear the critical line as in the case of $Q\neq 0$. 
Note also, that in general HD gravity the structure of Hawking-Page phase 
transitions is getting also quite complicated \cite{noplb}.
It would be interesting to investigate the role of higher derivative 
terms in charged AdS BHs considered in refs.\cite{beh,cvetic}.
As the final remark we note that the sign of BH entropy should be checked 
for all phases in the above phase diagrams. That will be done in section 
eight.

\section{Higher derivative electromagnetic terms\label{Sec3}}

In the string theory, $R^2$-terms appear as 
$\alpha'$-corrections in the low energy effective action.
In other words, there is good motivation to study higher derivative gravity.
However, for some versions of superstring there appear also gauge fields 
as well as 
 higher order terms for the gauge fields.
Hence, the natural question maybe about existence of charged AdS BHs 
in the presence of such higher derivative gauge terms.
Let us show that it is possible.
 As an extension of the above 
case where the usual $F^2$ action couples with $R^2$ gravity, 
we add $F^4$-terms to the matter action in (\ref{GBxiii})
\bea
\label{GBivxviii}
&& S_{\rm matter}= -{1 \over 4g^2}\int d^{d+1}x \sqrt{-g} 
g^{\mu\nu}g^{\rho\sigma} F_{\mu\rho} F_{\nu\sigma} \\
&& \quad + \int d^{d+1}x \sqrt{-g} \left(
\alpha g^{\mu\nu}g^{\rho\sigma} g^{\eta\tau}g^{\zeta\xi} 
+ \beta g^{\xi\mu}g^{\rho\nu} g^{\sigma\eta}g^{\zeta\tau} 
\right)
F_{\mu\rho} F_{\nu\sigma} F_{\eta\zeta}F_{\tau\xi}\ .\nonumber
\eea
Then the energy-momentum tensor is given by
\bea
\label{GBivxix}
T^{\mu\nu}&=&-{1 \over 4g^2}\left({1 \over 2}g^{\mu\nu} 
F_{\rho\sigma} F^{\rho\sigma} 
 - 2 F^\mu_{\ \sigma}F^{\nu\sigma}\right) \nn
&&  +\alpha \left\{{1 \over 2}g^{\mu\nu} 
\left(F_{\rho\sigma} F^{\rho\sigma}\right)^2 
 - 4 F^\mu_{\ \sigma}F^{\nu\sigma}F_{\rho\tau} F^{\rho\tau}
\right\} \nn
&& +\beta \left\{{1 \over 2}g^{\mu\nu} 
F_{\xi\rho} F^{\rho\sigma} F_{\sigma\zeta}F^{\zeta\xi}
 -4 F^{\mu\sigma} F_{\sigma\zeta}F^{\zeta\xi}
F_\xi^{\ \nu} \right\} \ ,
\eea
and by the variation over $A_\mu$, one has the following equation:
\be
\label{GBvx}
0=\partial_\nu\left\{\sqrt{-g}\left(-{1 \over g^2}F^{\nu\mu} 
+8\alpha F^{\nu\mu} \left(F_{\rho\sigma} 
F^{\rho\sigma}\right) +8\beta F_{\rho\sigma} F^{\nu\rho}
F^{\sigma\mu} \right)\right\}\ .
\ee
Assuming that $F_{tr}=-F_{rt}$ only depends on $r$ and other 
components in the field strength $F_{\mu\nu}$ vanish. 
Eq.(\ref{GBvx}) gives
\be
\label{GBvxi}
0={d \over dr}\left\{{1 \over g^2}r^3
\e^{-\left(\nu+\lambda\right)}F_{tr} 
+ 8\left(2\alpha + \beta\right)r^3\e^{-3\left(\nu+\lambda
\right)}F_{tr}^3
\right\}\ , 
\ee
that is, 
\be
\label{GBvxii}
{1 \over g^2}r^3
\e^{-\left(\nu+\lambda\right)}F_{tr} 
+ 8\left(2\alpha + \beta\right)\e^{-3\left(\nu+\lambda
\right)}F_{tr}^3 ={Q \over r^3}\ .
\ee
Here $Q$ is a constant of the integration. Eq.(\ref{GBvxii}) 
can be solved, by using $\zeta$ in (\ref{GBxxxib}), as
\bea
\label{GBvxiii}
F_{tr}&=&\e^{\nu+\lambda}f(r) \ ,\nn
f(r)&=&\beta_+ + \beta_- \ ,
\beta_+ \zeta + \beta_- \zeta^2 \ ,
\quad \beta_+ \zeta^2 + \beta_- \zeta \ ,\nn
\beta_\pm^3&=&{1 \over 16(2\alpha + \beta)r^3} 
\left\{1 \pm \sqrt{1 + {r^6 \over 
54(2\alpha + \beta)g^2}}\right\}\ .
\eea
Then the energy-momentum tensor is
\bea
\label{GBvxiv}
T^{tt}&=& \e^{-2\nu}\left\{ {1 \over 4g^2}f(r)^2 
+ 3\left(2\alpha + \beta\right)f(r)^4\right\} \ ,\nn
T^{rr}&=& - \e^{-2\lambda}\left\{ {1 \over 4g^2}f(r)^2 
+ 3\left(2\alpha + \beta\right)f(r)^4\right\} \ ,\nn
T^{ij}&=& \left\{ {1 \over 4g^2}f(r)^2 
+ \left(2\alpha + \beta\right)f(r)^4\right\}
{\tilde g^{ij} \over r^2}\ .
\eea
Since Eq.(\ref{GBx}) is satisfied, we have $\nu=-\lambda$ in 
(\ref{GBxi}). Then from Eq.(\ref{GBxii}) with $d=4$, we find
\bea
\label{GBvxv}
0&=&2c\left\{ - 2k \e^{-2\lambda} 
+ 2 \e^{-4\lambda}\right\} 
 - {3 \over \kappa^2} r^2\e^{-2\lambda} 
 + {3 \over 2}ck^2 + {3k \over 2\kappa^2}r^2 \nn
&& - {\Lambda \over 4}r^4 + {Q^2 \over g^2r^2} 
 - 2\int dr r^3 \left\{ {1 \over 4g^2}f(r)^2 
+ 3\left(2\alpha + \beta\right)f(r)^4\right\} \ .
\eea
This equation demonstrates that charged AdS BH maybe constructed 
even when higher derivative gauge field terms present.
All calculations of the previous section may be repeated but 
the obtained results are extremely complicated so we will not present 
more details on this.   

It would be also very interesting to consider 
the  FRW brane universe embedded into charged 5d AdS BH.
Recently in \cite{AM}, the brane dynamics in the context of the 
Randall-Sundrum model \cite{RS} has been discussed  in frames of 
Einstein-Gauss-Bonnet gravity. 
   However, the charge has not been taken 
into account. 
 The early time brane-world cosmology in our theory will be discussed 
elsewhere \cite{lidsey}.

\section{dS/CFT correspondence and localization of brane gravity}

AdS/CFT duality (for a review, see \cite{AdS}) relates 
quantum gravity on AdS $D$-dimensional space with 
boundary CFT living in one dimension less. In the recent works 
 \cite{strominger,ds1} (for further development, see \cite{ds3}) it has been 
suggested dS/CFT correspondence in the similar sense as above AdS/CFT
correspondence. The fact that quantum gravity in de Sitter space 
may have some holographic dual has been also mentioned 
in several papers \cite{ds2}. 
The reason why AdS/CFT can be expected is the isometry of 
$d+1$-dimensional anti-de Sitter space, which is $SO(d,2)$ 
symmetry. It is identical with the conformal symmetry of 
$d$-dimensional Minkowski space. We should note, however, the 
$d+1$-dimensional de Sitter space has the isometry of 
$SO(d+1,1)$ symmetry, which can be a conformal symmetry of 
$d$-dimensional Euclidean space. Then it might be natural to 
expect the correspondence between $d+1$-dimensional de Sitter 
space and $d$-dimensional euclidean conformal symmetry (dS/CFT 
correspondence\cite{strominger}). In fact, the metric of 
$D=d+1$-dimensional anti de Sitter space (AdS) is given by
\be
\label{AdSm}
ds_{\rm AdS}^2=dr^2 + \e^{2r}\left(-dt^2 + \sum_{i=1}^{d-1}
\left(dx^i\right)^2\right)\ .
\ee
In the above expression, the boundary of AdS lies at 
$r=\infty$. If one exchanges the radial coordinate $r$ 
and the time coordinate $t$, we obtain the metric of the 
de Sitter space (dS): 
\be
\label{dSm}
ds_{\rm dS}^2=-dt^2 + \e^{2t}\sum_{i=1}^d
\left(dx^i\right)^2\ .
\ee
Here $x^d=r$. Then there is a boundary at $t=\infty$, where the 
Euclidean conformal field theory (CFT) can live and one expects 
dS/CFT correspondence. Note that the patch given by (\ref{dSm}) 
only covers half of de Sitter space. Replacing $t$ by $-t$, 
we obtain a patch which covers the other half of the de Sitter 
space. 

Having such deep similarity between AdS and dS spaces (black holes)
 it is natural to extend the above discussion for dS black holes.
Before going to it let us give several remarks about localization of 
gravity 
on dS brane embedded into SdS bulk. This is the crucial property of 
gravity in AdS/CFT correspondence.

AdS$_5$/CFT$_4$ correspondence tells us that the effective 
action $W_{\rm CFT}$ of CFT in 4 dimensions is given by the path 
integral of the supergravity in 5 dimensional AdS space:
\bea
\label{A1}
\e^{-W_{\rm CFT}}&=&\int [dg][d\varphi]\e^{-S_{\rm grav}}, \\
S_{\rm grav}&=&\SEH + \SGH + S_1 + S_2 + \cdots, \nn
\SEH&=&{1 \over 16\pi G}\int d^5 x \sqrt{\gfv}\left(R_{(5)} 
 + {12 \over l^2} + \cdots \right), \nn
\SGH&=&{1 \over 8\pi G}\int d^4 x \sqrt{\gfr}\nabla_\mu n^\mu, \nn
S_1&=& -{1 \over 8\pi G}\int d^4 x \sqrt{\gfr}\left({3 \over l}
+ \cdots \right), \nn
S_2&=& -{l \over 16\pi G}\int d^4 x \sqrt{\gfr}\left(
{1 \over 2}R_{(4)} + \cdots \right), \nn
&& \cdots \ . \nonumber
\eea
Here $16\pi G=\kappa^2$ and $\varphi$ expresses the (matter) 
fields besides the graviton. 
$\SEH$ corresponds to the Einstein-Hilbert action and 
$\SGH$ to the Gibbons-Hawking surface counter term and 
$n^\mu$ is the unit vector normal to the boundary. 
$S_1$, $S_2$, $\cdots$ correspond to the surface counter terms, 
which cancell the divergences when the boundary in AdS$_5$ goes to 
the infinity. 

In \cite{HHR}, two 5 dimensional balls $B_5^{(1,2)}$ are 
glued on the boundary, which is 4 dimensional sphere $S_4$. Instead 
of $S_{\rm grav}$, if one considers the following action $S$
\be
\label{A2}
S=\SEH + \SGH + 2S_1=S_{\rm grav} + S_1 - S_2 - \cdots,
\ee
for two balls, using (\ref{A1}), one gets the following 
boundary theory in terms of the partition function \cite{HHR}:
\bea
\label{A3}
&& \int_{B_5^{(1)} + B_5^{(1)} +S_4} [dg][d\varphi]\e^{-S} \nn
&=& \left(\int_{B_5} [dg][d\varphi]\e^{-\SEH - \SGH - S_1} 
\right)^2 \nn
&=&\e^{2S_2 + \cdots}\left(\int_{B_5} [dg][d\varphi]
\e^{-S_{\rm grav}} \right)^2 \nn
&=&\e^{-2W_{\rm CFT}+2S_2 + \cdots}\ .
\eea
Since $S_2$ can be regarded as the Einstein-Hilbert action on 
the boundary, which is $S_4$ in the present case, the gravity 
on the boundary becomes dynamical. The 4 dimensional gravity 
is nothing but the gravity localized on the brane in the 
Randall-Sundrum model \cite{RS}. 

For ${\cal N}=4$ $SU(N)$ Yang-Mills theory, the AdS/CFT dual is 
given by identifying
\be
\label{AdSCFT}
l=g_{\rm YM}^{1 \over 2}N^{1 \over 4}l_s\ ,
\quad {l^3 \over G}={2N^2 \over \pi}\ .
\ee
Here $g_{\rm YM}$ is the coupling of the Yang-Mills theory and 
$l_s$ is the string length. Then (\ref{A3}) tells that the 
RS model is equivalent to  a CFT (${\cal N}=4$ $SU(N)$ 
Yang-Mills theory) coupled to 4 dimensional gravity including 
some correction coming from the higher order counter terms with 
a Newton constant given by
\be
\label{coupling}
G_4=G/l\ .
\ee

This is an excellent explanation \cite{HHR} to why gravity 
is trapped on the brane.

Recently, in \cite{BBM}, it has been pointed out that even in de Sitter 
space, the bulk action diverges when we substitute the classical 
solution, which is the fluctuation around the de Sitter space 
in (\ref{dSm}) and we need counterterms again. The divergence 
occurs since the volume of the space diverges when $t\rightarrow 
\infty$ (or $t\rightarrow -\infty$ after replacing $t$ by $-t$ 
in another patch). Then one should put the surface counterterms on the 
space-like branes which lie at $t\rightarrow \pm\infty$. 
Then dS/CFT correspondence should be given by, instead of 
(\ref{A1})
\bea
\label{A1b}
\e^{-W_{\rm CFT}}&=&\int [dg][d\varphi]\e^{-S_{\rm dS\,grav}}, \\
S_{\rm dS\,grav}&=&\SEH + \SGH + S_1 + S_2 + \cdots, \nn
\SEH&=&{1 \over 16\pi G}\int d^5 x \sqrt{-\gfv}\left(R_{(5)} 
 - {12 \over l^2} + \cdots \right), \nn
\SGH&=&{1 \over 8\pi G}\int_{M_4^+ + M_4^-} d^4 x 
\sqrt{\gfr}\nabla_\mu n^\mu, \nn
S_1&=& {3 \over 8\pi G l}\int_{M_4^+ + M_4^-} d^4 x 
\sqrt{\gfr}, \nn
S_2&=& {l \over 32\pi G}\int_{M_4^+ + M_4^-} d^4 x 
\sqrt{\gfr}\left(R_{(4)} + \cdots \right), \nn
&& \cdots \ . \nonumber
\eea
Here $S_1$, $S_2$, $\cdots$ correspond to the surface counter terms, 
which cancell the divergences in the bulk action and $M_4^\pm$ 
expresses the boundary at $t\rightarrow \pm \infty$.  

Let us consider two copies of the de Sitter spaces dS$_{(1)}$ 
and dS$_{(2)}$. We also put one or two of the space-like 
branes, which can be 
regarded as boundaries  connecting  two bulk de Sitter 
spaces, at finite $t$. Then if one considers the 
following action $S$ instead of $S_{\rm dS\,grav}$ in 
(\ref{A1b}), 
\be
\label{A2b}
S=\SEH + \SGH + 2S_1=S_{\rm dS\,grav} + S_1 - S_2 - \cdots,
\ee
one obtains the following boundary theory in terms of 
the partition function:
\bea
\label{A3b}
&& \int_{{\rm dS_5^{(1)} + dS_5^{(1)} +M_4^+ + M_4^-}} 
[dg][d\varphi]\e^{-S} \nn
&=& \left(\int_{dS_5} [dg][d\varphi]\e^{-\SEH - \SGH - S_1} 
\right)^2 \nn
&=&\e^{2S_2 + \cdots}\left(\int_{dS_5} [dg][d\varphi]
\e^{-S_{\rm grav}} \right)^2 \nn
&=&\e^{-2W_{\rm CFT}+2S_2 + \cdots}\ .
\eea
Since $S_2$ can be regarded as the Einstein-Hilbert action on 
the boundary, the gravity on the boundary becomes dynamical 
again. 

Hence, we demonstrated that in frames of dS/CFT correspondence 
the gravity trapping on the brane embedded into dS bulk occurs 
in the similar way as in Randall-Sundrum scenario. It would be 
interesting to give also the direct proof of this property,
using gravitational perturbations.

\section{Thermodynamics of de Sitter space for Einstein gravity}

$D$-dimensional de Sitter space can be realized by embedding 
it into the flat $(D+1)$-dimensional spacetime, whose metric is 
given by 
\be
\label{i}
ds^2=-\left(dX^0\right)^2 + \sum_{i=1}^D \left(dX^i\right)^2\ .
\ee
Here $X^\mu$ ($\mu=0,1,2,\cdots,D$) are orthogonal coordinates 
in $(D+1)$ dimensional spacetime. $D$-dimensional de Sitter space 
is the surface given by a constraint
\be
\label{ii}
l^2=-\left(X^0\right)^2 + \sum_{i=1}^D \left(X^i\right)^2\ .
\ee
We now choose unconstrained coordinates by defining
\be
\label{iii}
X^D + X^0 =\e^{\tau \over l}\ ,\quad X^i = 
\e^{\tau \over l}x^i\ (i=1,2,\cdots,D-1)
\ee
and by solving $X^D-X^0$ with respect to $t$ and $x^i$:
\be
\label{iv}
X^D - X^0 = l^2\e^{-{\tau \over l}} - \e^{\tau \over l}
\sum_{i=1}^{D-1} \left(x^i\right)^2\ .
\ee
Then the metric (\ref{i}) induces the following metric of 
the de Sitter space:
\be
\label{v}
ds_{\rm dS1}^2=- d\tau^2 + \e^{2\tau \over l}
\sum_{i=1}^{D-1} \left(dx^i\right)^2\ .
\ee
There are several expressions for the de Sitter space. 
Instead of (\ref{iii}), we now choose the polar (spherical) 
coordinates for $X^i$ ($i=1,2,\cdots, D-1$) as follows
\be
\label{viA} 
\sum_{i=1}^{D-1} \left(dX^i\right)^2= dr^2 
+ r^2 d\Omega_{D-2}^2\ .
\ee
Here $d\Omega_{D-2}^2$ is the metric of the ($D-2$)-dimensional 
sphere and the radial coordinate $r$ is given by
\be
\label{vii}
r^2=\sum_{i=1}^{D-1} \left(X^i\right)^2\ .
\ee
One can choose
\be
\label{viii}
X^D+X^0=\sqrt{l^2 - r^2}\e^{t \over l}\ ,\quad 
X^D-X^0=\sqrt{l^2 - r^2}\e^{-{t \over l}}
\ee
if $l^2>r^2$ or 
\be
\label{ix}
X^D+X^0=\sqrt{r^2 - l^2}\e^{t \over l}\ ,\quad 
X^D-X^0=-\sqrt{r^2 - l^2}\e^{-{t \over l}}
\ee
if $l^2 < r^2$. 
Eqs.(\ref{vii}) and (\ref{viii}) or (\ref{ix}) satisfy the 
constraint (\ref{ii}). Then the metric (\ref{i}) induces 
the following metric of the de Sitter space:
\be
\label{xA}
ds_{\rm dS2}^2=- \left(1-{r^2 \over l^2}\right)dt^2 
+ \left(1-{r^2 \over l^2}\right)^{-1}dr^2
+ r^2 d\Omega_{D-2}^2\ , 
\ee
which corresponds to the metric of Schwarzshild-de Sitter 
space with the balck hole mass $\mu=0$:
\be
\label{xAA}
ds_{\rm SdS}^2=- \left(1-{r^2 \over l^2} -\mu r^{2-d}
\right)dt^2 + \left(1-{r^2 \over l^2
 - \mu r^{2-d}}\right)^{-1}dr^2 + r^2 d\Omega_{D-2}^2\ . 
\ee
Using Eqs. (\ref{iii}), (\ref{vii}) and 
(\ref{viii}) or (\ref{ix}), one finds the relations between 
two coordinate systems  (\ref{v}) and (\ref{xAA}):
\be
\label{xi}
r^2 = \e^{2\tau \over l}\sum_{i=1}^{D-1} \left(x^i\right)^2
\ ,\quad \e^{2t \over l}=\pm {\e^{2\tau \over l} \over 
l^2 - \e^{2\tau \over l}\sum_{i=1}^{D-1} \left(x^i\right)^2}\ .
\ee
In the second equation of (\ref{xi}) the plus (minus) sign 
corresponds to the case of $l^2>r^2$ ($l^2<r^2$). 
In the coordinate system (\ref{v}), conformal field theory 
would exist on the brane at $t\rightarrow +\infty$ with 
fixed $x^i$. In the limit, the equations (\ref{xi}) have
the following forms:
\be
\label{xiA}
r^2 = \e^{2\tau \over l}\sum_{i=1}^{D-1} \left(x^i\right)^2
\ ,\quad \e^{2t \over l}\rightarrow 
{1 \over \sum_{i=1}^{D-1} \left(x^i\right)^2}\ .
\ee
Therefore the CFT brane corresponds to $r\rightarrow +\infty$ 
in the coodinate system in (\ref{xAA}). This tells that 
the CFT brane should exist outside the cosmological brane 
at $r=l$. 

Since there are two horizons in the Schwarzschild-de Sitter 
spacetime, 
there are two Hawking temperatures. Then if we put a brane into such 
the spacetime, what is the temperature measured on the brane?
Before going to the Gauss-Bonnet theory 
 we consider the 5-dimensional Einstein theory, where the 
Schwarzschild-de Sitter spacetime (\ref{xAA}) is given by
\be
\label{dSE1}
ds^2_{\rm Einstein}=-\e^{2\rho}dt^2 
+ \e^{-2\rho}dr^2 + r^2 d\Omega_3^2\ ,\quad 
\e^{2\rho}= 1 - {\mu \over r^2} - {r^2 \over l^2}\ .
\ee
Then there is a black hole horizon at
\be
\label{dSE2}
r^2=r^2_{\rm bh}\equiv {l^2 - \sqrt{l^2 - 4\mu l^2} \over 2}
\ee
and a cosmological horizon at 
\be
\label{dSE3}
r^2=r^2_{\rm csm}\equiv {l^2 + \sqrt{l^2 - 4\mu l^2} \over 2}\ .
\ee
Then the corresponding Hawking temperatures are given by
\be
\label{dSE4}
4\pi T_{\rm bh,csm}=\left|{2\mu \over r_{\rm bh,csm}^3} 
 - {2 r_{\rm bh,csm} \over l^2}\right|
= {2\sqrt{l^4 - 4\mu l^2} \over l^2 r_{\rm bh,csm}}\ .
\ee
Eq.(\ref{dSE4}) suggests that the temperature measured 
on the brane should be 
\be
\label{dSE4b}
T={r_{\rm bh} \over r}T_{\rm bh}
={r_{\rm csm} \over r}T_{\rm csm}\ .
\ee
In \cite{BBM}, by using the surface energy-momentum tensor, 
the mass $E$ of the Schwarzschild-de Sitter spacetime (in planar
coordinates) has 
been calculated:
\be
\label{dSE5}
E={3\pi l^2 \over 32 G}- {3\pi \mu \over 8G}\ ,
\ee
which can be regarded as the thermodynamical energy. 
Here $16\pi G=\kappa^2$.
As one sees the mass of  Nariai BH is zero (in four dimensions 
it becomes negative using similar formula)  while de Sitter space mass
is positive (it is zero in four dimensions). Note that recently there 
appeared very complete treatment of conserved BH mass in different 
dimensions\cite{mann}.

 Let us assume the entropy is given in terms of
 the horizon area
\be
\label{dSE6}
S_{\rm bh,csm}={\pi^2 r_{\rm bh,csm}^3 \over 4G}\ .
\ee
Here $S_{\rm bh}$ ($S_{\rm csm}$) corresponds to the 
black hole (cosmological) horizon. Then one gets 
\be
\label{dSE7}
{dS_{\rm bh,csm} \over dE}={{dS_{\rm bh,csm} \over d\mu}
\over {dE \over d\mu}}=\mp {1 \over T_{\rm bh,cms}}\ .
\ee
In (\ref{dSE7}), the minus (plus) sign corresponds to the 
black hole (cosmological) horizon. Then the entropy 
$S_{\rm csm}$ reproduces the usual thermodynamical relation 
${dS \over dE}={1 \over T}$. 
This indicates that the brane should be outside the 
cosmological horizon, as in the CFT boundaries, and only the 
cosmological horizon can be observed from the brane while the 
black hole horizon cannot be observed.  
 
 The expression for the energy (mass) of 3d SdS BH obtained in 
\cite{Myung} follows the expression in \cite{BBM} and 
so-called black hole mass parameter, which corresponds to 
$\mu$ in (\ref{dSE5}), enters in the expression with the 
minus sign as in (\ref{dSE5}). In 3d Shwarzschild-de Sitter 
black hole, there does not appear the black hole horizon and the 
horizon appeared there corresponds to the cosmological one as 
the radius is finite even in the limit of $\mu=0$. Then 
the Hawking temperature  is unique and it 
corresponds to the cosmological one. Then the results in 
\cite{Myung} are consistent with our Eqs. 
(\ref{dSE4b})-(\ref{dSE7}). From the viewpoint 
of dS/CFT, the entropy seems to correspond to the cosmological one. 
This is probably because the CFT brane is spacelike and it should 
lie outside the cosmological horizon. Nevertheless, this does not mean the
dynamical brane, as Schwarzschild-AdS in \cite{SV}, must be 
outside the cosmological horizon when one studies 
the brane FRW-like equation. 

\section{Mass of SdS Black Hole in Gauss-Bonnet gravity}

After above remarks on the properties of SdS BHs in Einstein gravity,
one can account for the modifications which occur
in Gauss-Bonnet gravity.

When $\Lambda$ is positive or $c$ is negative, we can find 
$l^2$  (\ref{GBxxvi}) can be formally negative. Then the spacetime 
can be asymptotically de Sitter. By replacing $l^2$ in 
(\ref{GBxxvi}) with $-l^2$:
\be
\label{dSi}
{1 \over l^2}= -{1 \over 4c\kappa^2 }\left(1\pm 
\sqrt{1 + {2c\Lambda \kappa^4 \over 3}}\right) \ ,
\ee
instead of (\ref{GBxxviii}), we obtain
\bea
\label{dSii}
\e^{2\nu}&=&\e^{-2\lambda} \nn
&=& {1 \over 2c}\left\{2c + {r^2 \over 2\kappa^2 } \right.\nn
&& \left. \pm \sqrt{ {r^4 \over 4\kappa^4}
\left({4c\kappa^2 \over l^2} +1
\right)^2 + {2c\mu \over \kappa^2}\left({4c\kappa^2 \over l^2} +1
\right) - {2cQ^2 \over 3g^2r^2 }} \right\}\ ,\\
\label{dSiii}
0 &=& r_H^6 - {1 \over \kappa^2
\left({2c \over l^4} + {1 \over \kappa^2 
l^2}\right)}r_H^4 + { \mu \left( 
{6c \over l^2} + {3 \over 2\kappa^2}\right) \over 
{2c \over l^4} + {1 \over \kappa^2 l^2}}r_H^2 \nn
&& - {Q^2 \over 3g^2 \left({2c \over l^4} + {1 \over \kappa^2 
l^2}\right)}\ ,\\\
\label{dSiv}
4\pi T_H &=& \left.\left(\e^{2\nu}\right)'\right|_{r=r_H} \nn
&=& {1 \over 2}\left(2c + {r_H^2 \over 2\kappa^2 }
\right)^{-1}\left[ {2r_H \over \kappa^2 } 
 - {8c r_H^3 \over l^4} - {4 r_H^3 \over \kappa^2 l^2}
 - {2 Q^2 \over 3 g^2 r_H^3} \right]\ .
\eea
Here we put $k=2$ since the horizon should be 3d sphere in the 
de Sitter background. 
In $R^2$-gravity, whose action is given by (\ref{vi}), the 
effective coupling constant can be given by the following 
replacement 
\cite{NOOr2a}
\be
\label{dSv}
{1 \over 16\pi G}= {1 \over \kappa^2}
\rightarrow {1 \over \kappa^2} + {40a \over l^2} 
 + {8b \over l^2} + {4c \over l^2}\ .
\ee
In case of the Gauss-Bonnet theory ($a=c$ and $b=-4c$), 
we obtain 
\be
\label{dSvi}
{1 \over 16\pi G}= {1 \over \kappa^2}
\rightarrow {1 \over \kappa^2} + {12c \over l^2}\ .
\ee
Since the asymptotic behavior is de Sitter, we may conjecture to be able
to 
use the formula in (\ref{dSE5}) and obtain the expression of 
the mass
\be
\label{dSvii}
E=\left({1 \over \kappa^2} + {12c \over l^2}\right)
\left({3\pi^2 l^2 \over 2}- 6\pi^2 \mu \right)\ .
\ee
Furthermore by using (\ref{dSi}) for $l^2$ and (\ref{GBxxvi}) 
for $\mu$, one gets
\be
\label{dSviii}
E=
{1 \over \kappa^2 }\left(4\pm 
3\sqrt{1 + {2c\Lambda \kappa^4 \over 3}}\right) 
\left( {6\pi^2  \over \Lambda\kappa^2}\left(1\mp 
\sqrt{1 + {2c\Lambda \kappa^4 \over 3}}\right)
\mp {2\pi^2 \kappa^2 \tilde C \over 
\sqrt{1 + {2c\Lambda \kappa^4 \over 3}}}\right)\ .
\ee
In the next section we will prove this formula (actually, its more general
variant for HD gravity) using the surface
counterterms in 
the $R^2$-gravity \cite{NOOr2a,HDf}. The energy 
(\ref{dSvii}) is zero when ${1 \over \kappa^2} + {12c \over l^2}=0$ 
or ${\pi^2 l^2 \over 2}- 2\pi^2 \mu =0$. The latter 
corresponds to the Nariai space as in \cite{BBM}, which tells 
that the Nariai space is real ground state whose energy is lower 
than the pure de Sitter spacetime.
Nevertheless, as one sees at the critical point BH mass disappears for any 
SdS BH in GB gravity. This is probably related with the fact that HD
gravity in general may not respect strong energy condition 
what results in the famous unitarity problem in four dimensions.
 The phenomena similar to the 
former case ${1 \over \kappa^2} + {12c \over l^2}=0$ occurs in 
Schwarzschild-AdS spacetime \cite{NOOr2a}, where the classical 
action vanishes then at the critical point
\be
\label{cr1}
{c\kappa^2 \over l^2}=-{1 \over 12}\ .
\ee
Then at the critical point, the black holes could be generated.

Even if the charge of the black hole vanishes, the 
Schwarzschild-de Sitter solution has two horizons, the black 
hole horizon and the cosmological one. The limit where the 
radius of the black hole horizon coincides with that of the 
cosmological one is called Nariai limit. We now consider the 
corresponding limit in the solution of (\ref{dSii}). For 
simplicity, we consider the $Q=0$ case. Then Eq.(\ref{dSiii}) 
can be easily solved and we find the radii of the two 
horizons are given by
\be
\label{dSix}
r_H^2=\left(r_H^\pm\right)^2 \equiv {1 \pm \sqrt{1 - 4\kappa^2\mu
\left( {2c \over 
l^4} + {1 \over \kappa^2 l^2}\right)
\left( {6c \over l^2} + {3 \over 2\kappa^2}\right) } 
\over 2\kappa^2 \left({2c \over l^4} + {1 \over \kappa^2 l^2}
\right)}\ .
\ee
Then the limit corresponding to the Nariai limit, where
\be
\label{dSx}
\left(r_H^+\right)^2 = \left(r_H^-\right)^2 =      
{1 \over 2 \kappa^2 \left({2c \over l^4} + {1 \over \kappa^2 l^2}
\right)}\ ,
\ee
appears when
\be
\label{dSxi}
\mu\rightarrow \mu_c\equiv 
{1 \over 4\kappa^2\left( {6c \over l^2} 
+ {3 \over 2\kappa^2}\right)
\left({2c \over l^4} + {1 \over \kappa^2 l^2}
\right)}\ .
\ee
Then the Hawking temperature $T_H$ (\ref{dSiv}) vanishes, 
what is consistent since even in the Einstein gravity, 
the Hawking temperature vanishes in the Nariai limit. 
One should also note that the energy $E$ (\ref{dSvii}) does 
not vanish when we substitute $\mu$ (\ref{dSxi}) into this  
expression.

\section{Surface counterterms in higher derivative gravity on 
SAdS and SdS backgrounds}

Let us discuss SAdS and SdS BHs in HD gravity using surface 
counterterms found in \cite{HDf,NOOr2a}. 
We start from the general $R^2$ part of the total action 
of $d+1$ dimensional $R^2$-gravity:
\be
\label{ai}
S_{R^2}=\int d^{d+1} x \sqrt{-g}\left\{a R^2 + b R_{\mu\nu} R^{\mu\nu} 
+ c  R_{\mu\nu\xi\sigma} R^{\mu\nu\xi\sigma}\right\}\ .
\ee
By introducing the auxiliary fields $A$, $B_{\mu\nu}$, 
and $C_{\mu\nu\rho\sigma}$, one can rewrite the action  
(\ref{ai}) in the following form:
\bea
\label{iia}
\tilde S_{R^2}&=&\int d^{d+1} x \sqrt{- g}
\left\{a \left(2 A R - A^2\right) + b \left(2 B_{\mu\nu} R^{\mu\nu}
 - B_{\mu\nu} B^{\mu\nu}\right) \right. \nn
&& \left. + c \left(2 C_{\mu\nu\xi\sigma} R^{\mu\nu\xi\sigma} - 
C_{\mu\nu\xi\sigma} C^{\mu\nu\xi\sigma}\right)\right\}\ .
\eea
Using the equation of the motion
\be
\label{iib}
A=R\ ,\quad B_{\mu\nu}= R_{\mu\nu}\ ,\quad 
C_{\mu\nu\rho\sigma}= R_{\mu\nu\rho\sigma}\ , 
\ee
we find the action (\ref{iia}) is equivalent to (\ref{ai}). 
Let us impose a Dirichlet type boundary condition, which 
is consistent with (\ref{iib}), 
$A=\left. R\right|_{\rm at\ the\ boundary}$, 
$B_{\mu\nu}=\left.t R_{\mu\nu}\right|_{\rm at\ the\ boundary}$, 
and $C_{\mu\nu\rho\sigma}=\left. R_{\mu\nu\rho\sigma}
\right|_{\rm at\ the\ boundary}$ and 
$\delta A = \delta B_{\mu\nu} = \delta C_{\mu\nu\rho\sigma}=0$ 
on the boundary. However, the conditions for $B_{\mu\nu}$ and 
$C_{\mu\nu\rho\sigma}$ are, in general, inconsistent. 
For example, even if $\delta B_{\mu\nu}=0$, we have 
$\delta B_{\mu}^{\ \nu}=\delta g^{\nu\rho}B_{\mu\rho}\neq 0$. 
Then one can impose boundary conditions on the scalar 
quantities:
\be
\label{bc1}
A= B_\mu^{\ \mu}= C^{\mu\ \nu}_{\ \mu\ \nu}= R\ ,
\quad n^\mu n^\nu B_{\mu\nu} 
= n^\mu n^\nu C_{\mu\rho\nu}^{\ \ \ \rho}= n^\mu n^\nu R_{\mu\nu} \ .
\ee
and 
\be
\label{bc2}
\delta A=\delta\left( B_\mu^{\ \mu}\right)
=\delta \left( C^{\mu\ \nu}_{\ \mu\ \nu}\right)
=\delta\left( n^\mu n^\nu B_{\mu\nu}\right) 
=\delta\left( n^\mu n^\nu C_{\mu\rho\nu}^{\ \ \ \rho}\right)=0\ .
\ee
Here $n^\mu$ is a unit vector perpendicular to the boundary. 
In order to realize the above boundary conditions (\ref{bc2}), 
we divide $ B_{\mu\nu}$ and $ C_{\mu\nu\rho\sigma}$ as 
follows:
\bea
\label{BCa1}
B_{\mu\nu}&=&\tilde B_{\mu\nu}+{1 \over d}\left(B_1 - B_2\right)
g_{\mu\nu}  - {1 \over d}\left(B_1 - (d+1)B_2\right)n_\mu n_\nu  \nn
C_{\mu\nu\rho\sigma}&=&\tilde C_{\mu\nu\rho\sigma} 
+ {1 \over d(d-1)}\left(C_1 - 2C_2\right)
\left(g_{\mu\rho} g_{\nu\sigma} - g_{\mu\sigma}g_{\nu\rho}
\right) \nn
&& - {1 \over d(d-1)} \left(C_1 - (d+1)C_2 \right)
\left(n_\mu n_\rho g_{\nu\sigma} 
+ n_\nu n_\sigma g_{\mu\rho} \right. \nn
&& \left.- n_\mu n_\sigma g_{\nu\rho}
 - n_\nu n_\rho g_{\mu\sigma}\right)\ .
\eea
Here 
\be
\label{BCa2}
B_1\equiv B_\mu^{\ \mu}\ ,\quad B_2\equiv n^\mu n^\nu B_{\mu\nu} 
\ ,\quad C_1\equiv C^{\mu\ \nu}_{\ \mu\ \nu}\ ,\mbox{and}, 
C_2\equiv  n^\mu n^\nu C_{\mu\rho\nu}^{\ \ \ \rho} \ ,
\ee
and $\tilde B_{\mu\nu}$ and $\tilde C_{\mu\nu\rho\sigma}$ are 
defined by (\ref{BCa2}) and satisfy the following equations:
\be
\label{BCa2b}
\tilde B_\mu^{\ \mu}= n^\mu n^\nu \tilde B_{\mu\nu} =0
\ ,\quad \tilde C^{\mu\ \nu}_{\ \mu\ \nu}= 
n^\mu n^\nu \tilde C_{\mu\rho\nu}^{\ \ \ \rho} =0\ .
\ee
If there appear $\tilde B_{\mu\nu}$ and 
$\tilde C_{\mu\nu\rho\sigma}$ in the final expressions, 
there are some ambiguities in the expression. 

Using the  conventions of curvatures in (\ref{curv}), one can 
further rewrite the action (\ref{iia}) in the following 
form:
\bea
\label{iiia}
\tilde S_{R^2}&=&2 \int_{{\rm surface}}d^d x 
\sqrt{-g_{(d)}}\left(
- \Gamma^\lambda_{\mu\rho}n_\nu + \Gamma^\lambda_{\mu\nu}n_\rho
\right) \nn
&& \times \left(a {\delta^\rho}_\lambda g^{\mu\nu} A 
+ b {\delta^\rho}_\lambda B^{\mu\nu} + c {C_\lambda}^{\ \mu\rho\nu}
\right)\nn
&& + \int d^{d+1}\left[\cdots\right]\ .
\eea
Here  $g_{(d)mn}$ is the $d$-dimensional boundary metric induced 
by $g_{\mu\nu}$. 
Now the bulk part of the action denoted by $[\cdots]$ 
does not contain the second order derivative of $g_{\mu\nu}$. 
Then the variational principle becomes well-defined if 
we add the following boundary term to the Einstein action: 
\bea
\label{iva}
\tilde S_{{\rm bndry}}&=&-2 \int_{{\rm surface}}d^d x 
\sqrt{-g_{(4)}}\left(- \Gamma^\lambda_{\mu\nu}n_\nu 
+ \Gamma^\lambda_{\mu\nu}n_\rho
\right) \nn
&&\times \left(a {\delta^\rho}_\lambda g^{\mu\nu} A 
+ b {\delta^\rho}_\lambda B^{\mu\nu} 
+ c {\hat C_\lambda}^{\mu\rho\nu} \right)\ .
\eea
The action (\ref{iva}) breaks the general covariance. 
We should note, however, that
\be
\label{E4}
\nabla_\mu n_\nu=\partial_\mu n_\nu
 - \Gamma_{\mu\nu}^\lambda n_\lambda \ ,\quad 
\nabla_\mu n^\nu=\partial_\mu n^\nu
 + \Gamma_{\mu\lambda}^\nu n^\lambda \ .
\ee
Then at least for the following metric
\be
\label{E7}
ds^2 = \left(1 + {\cal O}(y^2)\right)dy^2 
+ \hat g_{mn}(y,x^m) dx^m dx^n\ .
\ee
one can write the boundary action (\ref{iva}) as 
\bea
\label{va}
\hat S_{R^2\ {\rm bndry}}&=&\int d^dx \sqrt{-g_{(4)}}\left[
4a\nabla_\mu n^\mu A + 2b\left( n_\mu n_\nu \nabla^\lambda 
n_\lambda + \nabla_\mu n_\nu\right) B^{\mu\nu} \right. \nn
&& \left. + 4cn_\sigma n_\rho \nabla_\mu n_\nu C^{\sigma\mu\rho\nu}
\right]\ .
\eea

Before going forward, we consider the variation of $n_\mu$ and 
$\nabla_\mu n_\nu$. Let us assume the boundary or the brane 
is given by a scalar function by
\be
\label{n1}
f(x^\mu)=0\ .
\ee
Then on the boundary, we have $\partial_\mu f dx^\mu =0$, that is, 
the vector $\partial_\mu f$ is perpendicular to the surface, 
that is, $n_\mu\propto \partial_\mu f$. Since $n_\mu= n^\mu =1$, 
we find 
\be
\label{n2}
n_\mu ={\partial_\mu f \over \sqrt{g^{\rho\sigma}\partial_\rho f
\partial_\sigma f}}\ .
\ee
Then under the variation over $g_{\mu\nu}$, one gets
\be
\label{n3}
\delta n_\mu = {\partial^\xi f \partial^\zeta f 
\delta g_{\xi\zeta} 
\partial_\mu f \over \left(g^{\rho\sigma}\partial_\rho f 
\partial_\sigma f \right)^{3 \over 2}}
={1 \over 2}n_\mu n^\xi n^\zeta \delta g_{\xi\zeta}\ .
\ee
Then, since $\delta \Gamma^\kappa_{\mu\nu}= {1 \over 2} 
g^{\kappa\lambda} \left(
\nabla_\mu \delta g_{\nu\lambda} + \nabla_\nu 
\delta g_{\mu\lambda} - \nabla_\lambda \delta g_{\mu\nu}\right)$, 
we have
\be
\label{n4}
\delta\left(\nabla_\nu n_\mu\right) = \nabla_\nu\left(
n_\mu n^\xi n^\zeta \delta g_{\xi\zeta}\right) 
 - {1 \over 2} n^\lambda \left(
\nabla_\mu \delta g_{\nu\lambda} + \nabla_\nu 
\delta g_{\mu\lambda} - \nabla_\lambda \delta g_{\mu\nu}\right)\ .
\ee

Choosing $d+1=5$, we start again with the following bulk action:
\bea
\label{vib}
S&=&\int d^5 x \sqrt{-g}\left\{a \left(2 A R - A^2\right) 
+ b \left(2 B_{\mu\nu} R^{\mu\nu} - B_{\mu\nu} B^{\mu\nu}\right) 
\right. \nn
&& \left. + c \left(2 C_{\mu\nu\xi\sigma} R^{\mu\nu\xi\sigma}
 - C_{\mu\nu\xi\sigma} C^{\mu\nu\xi\sigma}\right)
+ {1 \over \kappa^2} R - \Lambda \right\}\ .
\eea
We also add the surface terms $S_b^{(1)}$ corresponding to 
Gibbons-Hawking surface term and  (\ref{va}) as well as $S_b^{(2)}$ 
which is the leading 
counterterm corresponding to the vacuum energy on the brane:
\bea
\label{Iiv}
S_b&=&S_b^{(1)} + S_b^{(2)} \nn
S_b^{(1)} &=& \int d^4 x \sqrt{- g}\left[
4 a \nabla_\mu n^\mu A + 2 b\left(n_\mu n_\nu \nabla_\sigma n^\sigma
 + \nabla_\mu n_\nu \right) B^{\mu\nu} \right. \nn
&& \left. + 8 c n_\mu n_\nu \nabla_\tau n_\sigma C^{\mu\tau\nu\sigma} 
+ {2 \over \kappa^2}\nabla_\mu n^\mu \right] \nn
S_b^{(2)} &=& - \int d^4 x \sqrt{- g_{(4)}} \left(\eta_1 + \eta_2 
R_{(4)}\right)\ .
\eea
Here $\eta_1$ and $\eta_2$ are constants, which can be determined 
by the condition that the total action is finite. 

Under the variation of $\delta g_{\mu\nu}$, the variation of the 
bulk action (\ref{vib}) gives the following contribution on the 
boundary:
\bea
\label{ABC1}
\lefteqn{\left.\delta S\right|_{\rm{boundary}} = \int 
\sqrt{-g_{(4)}}\left[\left({1 \over \kappa^2} + 2aA\right)
\left(n^\mu\nabla^\nu \delta g_{\mu\nu} - n^\rho\nabla_\rho 
\left(g^{\mu\nu}\delta g_{\mu\nu}\right)\right) \right.} \nn
&& + 2a\left\{- \nabla_\mu A n^\nu \delta g_{\mu\nu} 
+ n^\rho \nabla_\rho A g^{\mu\nu}\delta g_{\mu\nu}\right\} \nn
&& + 2b \left\{B^{\rho\mu} n^\nu \nabla_\rho \delta g_{\mu\nu} 
 - {1 \over 2} n^\tau B^{\mu\nu} \nabla_\tau 
 \delta g_{\mu\nu} - {1 \over 2} n_\rho B^{\rho\sigma} \nabla_\sigma
\left(g^{\mu\nu}\delta g_{\mu\nu}\right) \right. \nn
&& \left.  - n_\rho\nabla^\nu B^{\rho\mu} \delta g_{\mu\nu} 
+ {1 \over 2} n_\rho\nabla^\rho B^{\mu\nu} \delta g_{\mu\nu} 
+ {1 \over 2} n_\rho \nabla_\sigma B^{\rho\sigma} 
\left(g^{\mu\nu}\delta g_{\mu\nu}\right) \right\} \nn
&& + c\left\{ C^{\mu\nu\rho\sigma}\left(n_\rho \nabla_\nu 
\delta g_{\sigma\mu} - n_\rho \nabla_\mu \delta g_{\sigma\nu} 
 -n_\sigma \nabla_\nu \delta_{\rho\mu} 
 + n_\sigma \nabla_\mu \delta g_{\rho\nu} \right) \right. \nn
&& -n_\nu \nabla_\rho C^{\mu\nu\rho\sigma} 
\delta g_{\sigma\mu}
+ n_\mu \nabla_\rho C^{\mu\nu\rho\sigma} \delta g_{\sigma\nu} \nn
&& \left.\left. 
+ n_\nu \nabla_\sigma C^{\mu\nu\rho\sigma} \delta g_{\rho\mu}
 - n_\mu \nabla_\sigma C^{\mu\nu\rho\sigma} \delta g_{\rho\nu}
\right\}\right]\ .
\eea
The contribution from the surface terms in (\ref{Iiv}) is 
given by 
\bea
\label{ABC2}
\lefteqn{\delta S_b = \int \sqrt{-g_{(4)}}\left[\left\{ {2 \over 
\kappa^2} + 4aA + {2b\left(B_1 + (d-1) B_2\right) \over d} 
\right.\right.} \nn
&& \left. + {8c C_2 \over d}\right\}\left\{
{1 \over 2}\nabla_\rho n^\rho \left(g^{mn}\delta g_{mn} 
 - \nabla^\mu n^\nu \delta g_{\mu\nu} \right.\right. \nn
&& \left. + {1 \over 2}\nabla_\nu \left(n^\nu n^\rho n^\sigma 
\delta g_{\rho\sigma}\right) - {1 \over 2}\left( 
2 \nabla^\mu \delta g_{\mu\lambda} - \nabla_\lambda \left(
g_{\mu\nu} \delta g_{\mu\nu}\right)\right)\right\} \nn 
&& + 2b \left\{{1 \over 2}\nabla^\rho n^\sigma 
\tilde B_{\rho\sigma} g^{mn}\delta g_{mn} 
 - \nabla^\mu n_\rho \tilde B^{\nu\rho}\delta g_{\mu\nu} 
 - \nabla_\rho n^\mu \tilde B^{\rho\nu} \delta g_{\mu\nu} \right.\nn
&& \left. + {1 \over 2}\nabla_\nu \left(n_\mu n^\rho n^\sigma 
\delta g_{\rho\sigma} \tilde B^{\nu\mu} - {1 \over 2}n^\lambda 
\left(2\nabla_\nu \delta g_{\mu\lambda} - \nabla_\lambda 
\delta g_{\nu\mu}\right)\tilde B^{\nu\mu} \right)\right\} \nn
&& + 8c \left\{ {1 \over 2}\nabla^\mu n^\nu 
\tilde C_{\rho\mu\sigma\nu} n^\rho n^\sigma g^{mn}\delta g_{mn}
 -\nabla^\mu n_\rho \tilde C^{\ \eta\ \rho}_{\tau\ \sigma}n^\tau 
 n^\sigma \delta g_{\mu\eta} \right. \nn
&& -\nabla_\rho n^\mu \tilde C^{\ \rho\ \nu}_{\tau\ \sigma}n^\tau 
 n^\sigma \delta g_{\mu\nu} 
 -\nabla^\mu n^\nu \tilde C^\tau_{\ \mu\sigma\nu}n^\rho 
 n^\sigma \delta g_{\tau\rho} 
 -\nabla^\mu n^\nu \tilde C^{\ \ \tau}_{\rho\mu\ \nu}n^\rho 
 n^\sigma \delta g_{\tau\sigma} \nn
&& + {1 \over 2} \tilde C^{\ \nu\ \mu}_{\eta\ \zeta}
\nabla_\nu\left(n_\mu n^\rho n^\sigma \delta g_{\rho\sigma}
n^\eta n^\zeta \right) + \nabla^\mu\nu 
\tilde C_{\rho\mu\sigma\nu}n^\rho n^\sigma n^\zeta 
n^\xi \delta_{\zeta\xi} \nn 
&& \left. - {1 \over 2}n^\lambda \left(\nabla_\nu 
\delta g_{\mu\lambda} + \nabla_\mu \delta g_{\nu\lambda} 
- \nabla_\lambda \delta g_{\nu\mu} \right)
\tilde C^{\ \nu\ \mu}_{\rho\ \sigma}n^\rho n^\sigma  
\right\} \nn
&& \left. - \left\{{\eta_1 \over 2}g^{mn} 
+ \eta_2 \left(-R_{(4)}^{mn}
+ {1 \over 2}g^{mn}\right)\right\}\delta g_{mn}\right]\ .
\eea
Here $m,n=i,j,t$, $g_{mn}$ is the metric induced on the boundary 
and we have used the decompositions of $B_{\mu\nu}$ and 
$C_{\mu\nu\rho\sigma}$ in (\ref{BCa1}) and we have assumed $B_1$, 
$B_2$, $C_1$, $C_2$ and also $\tilde B_{\mu\nu}$ and 
$\tilde C_{\mu\nu\rho\sigma}$. 

One now considers the black hole like solution in the asymptotically 
anti-de Sitter space, as in 
(\ref{GBxxiii}) or (\ref{GBxxviii}) with $Q^2=q^2=0$, whose 
asymptotic behavior when $r$ is large is given by
\be
\label{ABC3}
\e^{2\nu}=\e^{-2\lambda} 
= {r^2 \over l^2} + {k \over 2} - {\mu \over r^2} 
+ {\cal O}\left(r^{-4}\right)\ .
\ee
We also put the boundary where $r$ is constant and finite. 
One takes a limit of $r\rightarrow \infty$ later. 
Then 
\bea
\label{ABC4}
&& A=B_2=C_2=-{20 \over l^2}\ ,\quad B_2=C_2=-{4 \over l^2} \nn
&& \tilde C_{trtr}=-{3\mu \over r^2}\ ,\quad 
\tilde C_{irjr}=-{\mu \over r^2}\tilde g_{ij}\ .
\eea
Then if we define the surface energy momentum tensor by
\be
\label{ABC5}
T^{mn}=T^{mn}_1 + T^{mn}_2\ ,\quad 
T^{mn}_1\equiv 2\left.{\delta S \over \delta g^{mn}}
\right|_{\rm boundary}\ ,\quad 
T^{mn}_2 \equiv 2{\delta S_b \over \delta g^{mn}}\ ,
\ee
one gets
\bea
\label{ABC6}
T^{mn}_1&=&\left({1 \over \kappa^2} - {40a \over l^2}
 - {8b \over l^2} - {4c \over l^2}\right)\left(2g^{mn} 
\nabla_\mu n^\mu - 2\nabla^m n^n\right)\ ,\nn
T^{mn}_2&=&-\eta_1 g^{mn} - 2\eta_2\left({1 \over 2}
g^{mn}R_{(4)} - R_{(4)}^{mn}\right)\ .
\eea
Note that there are no contributions from $\tilde B_{\mu\nu}$ 
and $\tilde C_{|mu\nu\rho\sigma}$. 
Especially we have
\bea
\label{ABC7}
T^{tt}_1&=&\left(-{1 \over \kappa^2} + {40a \over l^2}
+ {8b \over l^2}+{4c \over l^2}\right){6 \over lr^2}
\left(1 - {kl^2 \over 4r^2} + {3k^2 l^2 + 16 \mu l^2 
\over 32 r^4}\right) + {\cal O}\left(r^{-8}\right)\ ,\nn
T^{tt}_2&=&{\eta_1 l^2 \over r^2}\left(1 - {kl^2 \over 2r^2}
+ {k^2l^4 + 4\mu l^2 \over 32 r^4}\right)
+{3\eta_2 \over 2r^4}\left(1 - {kl^2 \over 2r^2}\right)
+ {\cal O}\left(r^{-8}\right)\ .
\eea
Then if we choose 
\be
\label{ABC8}
\eta_1={6 \over l}\left({1 \over \kappa^2} - {40a \over l^2}
- {8b \over l^2} - {4c \over l^2}\right)\ ,\quad 
\eta_2=kl^2\left({1 \over \kappa^2} - {40a \over l^2}
- {8b \over l^2} - {4c \over l^2}\right)\ ,
\ee
one obtains
\be
\label{ABC9}
T^{tt}={3 \over 4lr^6}\left({1 \over \kappa^2} - {40a \over l^2}
- {8b \over l^2} - {4c \over l^2}\right)\ .
\ee
As in \cite{BK,BBM}, the mass $M$ can be evaluated at the 3d 
surface where $t$ is a constant:
\be
\label{ABC10}
M=\int dx^3\sqrt{\tilde g}r^3 \e^\nu T^{tt}\left(\xi_t\right)^2
\ .
\ee
Here $\zeta^\mu$ is a unit vector parallel with the time 
coordinate $t$ and therefore $\zeta^t=\e^{-\nu}$ 
($\zeta_t=\e^\nu$) and $\zeta^\mu=\zeta_\mu=0$ ($\mu\neq t$). 
Note $\sqrt{\tilde g}\e^\nu = \sqrt{\det\,g_{mn}}$.  
Then 
\be
\label{ABC11}
M={3l^2 \over 16}V_3 \left({1 \over \kappa^2} - {40a \over l^2}
 - {8b \over l^2} - {4c \over l^2}\right)\left(k^2  
+ {16\mu \over l^2}\right)
\ee
Here $V_3$ is a volume of 3d surface with unit radius. 
When the surface is 3d sphere one has $k=2$ and $V_3=2\pi^2$. 
Then the expression in (\ref{ABC11}) has the following form:
\be
\label{ABC12}
M_{k=2}= {3l^2 \over 4}V_3 \left({1 \over \kappa^2}
 - {40a \over l^2} - {8b \over l^2} - {4c \over l^2}\right)
 \left(1 + {4\mu \over l^2}\right)\ ,
\ee
which reproduces the previous results when $a=b=c=0$ \cite{BK}. 
When we consider 5d de Sitter 
space instead of AdS space, similar calculation leads to
\be
\label{ABC12b}
M_{k=2,\,{\rm dS}}= {3l^2\pi^2 \over 2} 
\left({1 \over \kappa^2} + {40a \over l^2}
+ {8b \over l^2} + {4c \over l^2}\right)\left(1 
- {4\mu \over l^2}\right)\ ,
\ee
which reproduces the naive conjecture in (\ref{dSvii}) for the 
Gauss-Bonnet case. 
When $k=0$ in (\ref{ABC11}),
\be
\label{ABC11b}
{M_{k=0} \over V_3}=3l^2 \left({1 \over \kappa^2} - {40a \over l^2}
 - {8b \over l^2} - {4c \over l^2}\right) {\mu \over l^2}\ .
\ee
We should note that $M$ vanishes in the critical case that 
\be
\label{ABC13}
{1 \over \kappa^2} - {40a \over l^2} - {8b \over l^2}
 - {4c \over l^2}=0\ ,
\ee
where enormous number of black holes might be produced since all 
the black holes become massless. 
Thus, using surface counterterms method we calculated the conserved mass 
for HD gravity which generalizes the corresponding Einstein theory result
for 5d SAdS and SdS BHs.

\section{The tale of negative entropy or de Sitter/Anti- de Sitter Black 
Holes phase transition?}

In this section, we study the relation between SdS and SAdS BHs 
based on entropy considerations.
Before going to the Gauss-Bonnet case, we consider $c=0$ 
case in (\ref{iv}) without matter $S_{\rm matter}=0$. 
When $c=0$, Schwarzschild-anti de Sitter space is 
an exact solution:
\bea
\label{SAdSA}
ds^2&=&\hat G_{\mu\nu}dx^\mu dx^\nu \nn
&=&-\e^{2\rho_0}dt^2 + \e^{-2\rho_0}dr^2 
+ r^2\sum_{i,j}^{d-1} g_{ij}dx^i dx^j\ ,\nn
\e^{2\rho}&=&{1 \over r^{d-2}}\left(-\mu + {kr^{d-2} \over d-2} 
+ {r^d \over l^2}\right)\ .
\eea
The curvatures have the following form:
\be
\label{cvA}
\hat R=-{d(d+1) \over l^2}\ ,\quad 
\hat R_{\mu\nu}= - {d \over l^2}\hat G_{\mu\nu}\ .
\ee
In (\ref{SAdSA}), $\mu$ is the parameter corresponding to mass 
and the scale parameter $l$ is given by solving the following 
equation:
\be
\label{llA}
0={d^2(d+1)(d-3) a \over l^4} + {d^2(d-3) b \over l^4} \nn
- {d(d-1) \over \kappa^2 l^2}-\Lambda\ .
\ee
We also assume $g_{ij}$ corresponds to the Einstein manifold, 
again. In the following we concentrate on the case of $d=4$. 

By using the method parallel with section \ref{Sec2}, 
we found the free energy has the following form:
\be
\label{freeA}
F= -{V_{3} \over 8}r_{H}^{2} \left( {r_{H}^{2} \over l^{2}}
 - {k \over 2} \right)\left( {8 \over \kappa^2} 
 - {320 a \over l^2} -{64 b \over l^2} \right) \; .
\ee
Then the entropy ${\cal S }= -{dF \over dT_H}$ and the 
thermodynamical energy $E = F+T{\cal S}$ can be obtained 
as follows:
\bea
\label{entA}
{\cal S }&=&4V_{3}\pi r_H^3 
\left( {1 \over \kappa^2}- {40 a \over l^2}
 -{8 b \over l^2} \right)\ ,\\
\label{enerA}
E&=& 3V_{3}\mu 
\left( {1 \over \kappa^2}- {40 a \over l^2}
 -{8 b \over l^2} \right)\ , 
\eea
This seems to indicate that the contribution from 
the $R^2$-terms can be absorbed into the redefinition:
\be
\label{B16A}
{1 \over \tilde\kappa^2}={1 \over \kappa^2} - {40a \over l^2} 
 - {8b \over l^2}\ ,
\ee
although this is not true for $c\neq 0$ case. 

One can also start from the expression for the mass $M$ 
(\ref{ABC11}) with $c=0$ as the thermodynamical energy $E$:
\be
\label{ABCE1}
E=3V_3 \left({1 \over \kappa^2} - {40a \over l^2}
 - {8b \over l^2} \right)\left({k^2l^2 \over 16} 
+ {\mu \over l^2}\right)
\ee
The expression of energy $E$ (\ref{ABCE1}) is different 
from that in (\ref{enerA}) by a first $\mu$-independent 
term, which  comes from the AdS background. 
Since the horizon is defined by $\e^{2\rho}=0$,  using 
(\ref{SAdSA}), one finds the parameter $\mu$ in terms of $r_H$
\be
\label{ABCE2}
\mu = {kr_H^2 \over 2} + {r_H^4 \over l^2}\ ,
\ee
and the Hawking temperature in the following form:
\bea
\label{ht1}
T_H = {(e^{2\rho})'|_{r=r_{H}} \over 4\pi}
= {k \over 4\pi r_{H}} +{r_{H} \over \pi l^2} \ .
\eea
Using the thermodynamical relation $d{\cal S}={dE \over T}$, 
we find 
\be
\label{ABGE3}
{\cal S}=\int {dE \over T_H} = \int dr_H {dE \over d\mu}
{d \mu \over d r_H}{1 \over T_H} = {V_{3}\pi r_H^3 \over 2}
\left( {8 \over \kappa^2}- {320 a \over l^2}
 -{64 b \over l^2} \right) + {\cal S}_0\ .
\ee
Here $S_0$ is a constant of the integration. Up to the constant 
$S_0$, the expression (\ref{ABGE3}) is identical with 
(\ref{entA}). 
We should note that the entropy ${\cal S}$ in (\ref{entA}) 
becomes negative, when 
\be
\label{EnS1}
{8 \over \kappa^2}- {320 a \over l^2}
 -{64 b \over l^2}<0.\ .
\ee
This is true even for the expression (\ref{ABGE3}) for the 
black hole with large radius $r_H$ since $S_0$ can be neglected 
for the large $r_H$. 

We now investigate in more details what happens when 
Eq.(\ref{EnS1}) is satisfied. First we should note $l^2$ is 
determined by (\ref{llA}), which has, in case of $d=4$, the 
following form:
\be
\label{llA4}
0={80 a + 16 b \over l^4} - {12 \over \kappa^2 l^2} - \Lambda\ ,
\ee
There are two real solutions for $l^2$ when 
\be
\label{lll1}
\left({6 \over \kappa^2}\right)^2 
+ \left(80a + 16b\right)\Lambda \geq 0\ .
\ee
and the solutions are given by
\be
\label{lll2}
{1 \over l^2}={{6 \over \kappa^2}\pm \sqrt{
\left({6 \over \kappa^2}\right)^2 + \left(80a + 16b\right)\Lambda} 
\over 80a + 16b}\ .
\ee
Suppose $\kappa^2>0$. Then if 
\be
\label{lll3}
\left(80a + 16b\right)\Lambda>0\ ,
\ee
one solution is positive but another is negative. 
Therefore there are both of the asymptotically AdS solution 
and asymptotically dS one. Let us denote the positive solution for $l^2$ by 
$l_{\rm AdS}^2$ and the negative one by $-l_{\rm dS}^2$:
\be
\label{lll4}
l^2=l_{\rm AdS}^2,\ -l_{\rm dS}^2\ ,\quad 
l_{\rm AdS}^2,\ l_{\rm dS}^2 >0\ .
\ee
Then when the asymptotically AdS solution is chosen, the entropy 
(\ref{ABGE3}) has the following form: 
\be
\label{lll5}
{\cal S}_{\rm AdS}= {V_{3}\pi r_H^3 \over 2}
\left( {8 \over \kappa^2}- {320 a + 64b \over l_{\rm AdS}^2}
\right) \ .
\ee
Here we have chosen ${\cal S}_0=0$. On the other hand, 
when the solution is asymptotically dS, the entropy 
(\ref{ABGE3}) has the following form: 
\be
\label{lll6}
{\cal S}_{\rm dS}= {V_{3}\pi r_H^3 \over 2}
\left( {8 \over \kappa^2} + {320 a + 64b \over l_{\rm dS}^2}
\right) \ .
\ee
When
\be
\label{lll7}
{8 \over \kappa^2}- {320 a + 64b \over l_{\rm AdS}^2}<0\ ,
\ee
the entropy ${\cal S}_{\rm AdS}$ (\ref{lll5}) is negative!

There are different points of view to this situation.
Naively, one can assume that above condition is just the equation to
remove the non-physical domain of theory parameters. However,
 it is difficult to justify such proposal.
Why for classical action on some specific background 
there are parameters values which are not permitted?
Moreover, the string/M-theory and its compactification would tell us what 
are the values of the theory parameters.

 From another side, one can conjecture that classical thermodynamics is 
not applied here 
and negative entropy simply indicates to new type of instability in
asymptotically AdS black hole physics.
 Indeed, when Eq.(\ref{lll7}) is satisfied, 
since $80a + 16b >0$ (same range of parameters!),
 the entropy ${\cal S}_{\rm dS}$  
(\ref{lll6}) for asymptotically dS solution is positive. In other words, 
maybe the asymptotically dS solution would be preferrable?

On the other hand, when 
\be
\label{lll8}
{8 \over \kappa^2}+ {320 a + 64b \over l_{\rm dS}^2}<0\ ,
\ee
the entropy ${\cal S}_{\rm dS}$ in (\ref{lll6}) is negative 
and the asymptotically dS solution is  instable (or does not exist).
(Again, one can say that above condition defines the admissible parameter 
values). In this case, since $80a + 16b <0$, the entropy 
${\cal S}_{\rm AdS}$ in (\ref{lll5}) for asymptotically 
AdS solution is positive and the asymptotically AdS solution 
would be preferrable. Expression for the AdS black hole mass in 
(\ref{enerA}) or (\ref{ABCE1}) tells that when 
${8 \over \kappa^2}- {320 a + 64b \over l_{\rm AdS}^2}=0$, the 
AdS black hole becomes massless then there would occur the 
condensation of the black holes, which would make the transition to 
the dS black hole. On the other hand, when ${8 \over \kappa^2} + 
{320 a + 64b \over l_{\rm dS}^2}=0$, the 
dS black hole becomes massless then there would occur the 
condensation of the black holes and the AdS black hole 
would be produced. 
Note that above state with zero entropy (and also zero free energy and zero
conserved BH mass) is very interesting. Perhaps, this is some new state 
of BHs. As we saw that is this state which defines the border between 
physical SAdS (SdS) BH with positive entropy and  SdS 
(SAdS) BH with negative entropy.

Hence, there appeared some indication that 
some new type of phase transition (or phase transmutation) between
SdS and SAdS BHs in higher derivative gravity occurs.
Unfortunately, we cannot suggest any dynamical formulation to 
describe explicitly such phase transition (it is definitely phase transition
not in standard thermodynamic sense).

The remark is in order. In principle the string/M-theory and its 
compactification should predict the values for theory parameters.
It may occur that for realistic compactification the above conditions are 
never satisfied (and the entropy is always positive). From another side,
once higher derivative terms dominate the thermodynamical expressions one 
should in principle worry about higher powers of such terms.

 Let us 
consider 
now the entropy for Gauss-Bonnet case. For this purpose, we use the 
thermodynamical relation $d{\cal S}={dE \over T}$. For 
the Gauss-Bonnet case, the energy (\ref{ABC11}) has the 
following form:
\be
\label{EnS2}
E=M={3l^2 \over 16}V_3 \left({1 \over \kappa^2} - {12c \over l^2}
\right)\left(k^2  + {16\mu \over l^2}\right)
\ee
Solving (\ref{GBxxix}) with respect to $\mu$, we have 
\be
\label{EnS3}
\mu = {1 \over 2l^2}\left(k\epsilon - {1 \over 2}\right)^{-1}
\left\{ ( 2\epsilon - 1) r_H^4 
 - kr_H^2 l^2 - {Q^2\kappa^2 l^2 \over 3g^2 r_H^2}\right\}\ .
\ee
Here $\epsilon$ is defined in (\ref{GBivxiv}). 
Then using (\ref{EnS2}), (\ref{EnS3}), and the expression 
of the Hawking temperature (\ref{GBxxx}), the entropy can be 
obtained as
\bea
\label{EnS4} 
\lefteqn{{\cal S}=\int {dE \over T_H} = \int dr_H {dE \over d\mu}
\left.{\partial \mu \over \partial r_H}\right|_{Q^2} 
{1 \over T_H} } \\
&& =  {V_3 \over \kappa^2}\left({1 - 12 \epsilon 
\over 1-4\epsilon} \right)\left(4\pi r_H^3 
+ 24 \epsilon k \pi r_H\right) + {\cal S}_0\ .
\eea
Here ${\cal S}_0$ is a constant of the integration, which 
could be chosen to be zero if we assume ${\cal S}=0$ when 
$r_H=0$.  
The expression $\left.{\partial \,\cdot\, \over \partial r_H}
\right|_{Q^2}$ 
means the partial derivative with respect to $r_H$ when $Q^2$ is 
fixed. In the integration in (\ref{EnS4}), $Q^2$ is fixed. 
When $\epsilon=0$ ($c=0$), the expression reproduces the 
standard result 
\be
\label{EnS5} 
{\cal S}\rightarrow {4\pi V_3 r_H^3 \over \kappa^2}\ .
\ee
The entropy (\ref{EnS4}) becomes negative (at least for the 
large black hole even if ${\cal S}_0\neq 0$) when 
\be
\label{EnS6}
{1 \over 12}<\epsilon<{1 \over 4}\ .
\ee
Therefore the unitarity might be broken in this region but 
it might be recovered when $\epsilon>{1 \over 4}$. In the 
phase diagram  \ref{Fig1} in Section \ref{Sec2}, 
the region in (\ref{EnS6}) seems to be unphysical. Even in case 
$\epsilon<0$ ($k=2$), the entropy becomes negative when
\be
\label{EnSS1}
r_H^2 < - 12 \epsilon \ ,
\ee
if ${\cal S}_0=0$. Then the small black hole might be 
unphysical. 

The fact discovered in this section-that entropy for S(A)dS BHS in
gravity with higher derivatives terms maybe easily done to be negative 
by the corresponding choice of parameters is quite remarkable.
It is  likely that thermodynamics for black holes with negative entropies
should be reconsidered. In this respect the indication to possibility of 
some new type phase transition between SdS and SAdS BHs via the state 
with zero entropy is quite interesting and should be further investigated. 

\section{Discussion}

In summary, we presented the extensive study of charged SAdS and SdS BH 
thermodynamics for Einstein-Gauss-Bonnet gravity with electromagnetic field.
The number of related questions (like gravity localization in dS/CFT 
correspondence, etc) are also 
discussed. The careful investigation of Hawking-Page phase transitions 
between SAdS BH and pure AdS space in Einstein-GB-Maxwell theory 
is done. The dependence of such phase transition from the coefficient of 
GB term and from the electric charge is studied. The corresponding 
phase diagrams are presented.

The investigation of SdS BHs in higher derivative gravity is very much 
connected with dS/CFT correspondence. We presented surface counterterms 
derivation for higher derivative gravity on SdS and SAdS spaces.
The review of SdS BH thermodynamics in Einstein gravity is done.
It helps in the derivation of SdS BH thermodynamics for Einstein-GB 
gravity (with electromagnetic field). The conserved BH mass for such 
theory is derived.

The interesting property of higher derivative gravity (including GB theory)
is the possibility for SdS (SAdS) BH entropy to be negative (or zero) for 
some 
range of coefficients of higher derivative terms. This is presumably related
with the known fact that strong energy condition in higher derivative gravity
maybe violated. The simplest resolution of this phenomenon is to exclude 
the corresponding parameter values as non-physical ones. However, this 
may not be the good solution, as the definition of such (non-physical)
parameters region is very much dependent on the background space under 
consideration. Simply, for same version of higher derivative gravity 
say, SdS BH entropy is negative but SAdS BH entropy is positive and 
vice-versa. 
Moreover, it is clear that the effect of entropy sign change is an 
artifact of higher derivative terms competing with leading Einstein term.
This may not be the case in superstring theory.
Hence, our results hold within Einstein-Gauss-Bonnet theory.
However, one may point out the potential difficulties to embed this 
scenario in fully realistic M-/string theory compactification.

We expect that there occurs some new type of phase transition
between SdS and SAdS BHs: what was SAdS BH with negative entropy 
becomes SdS BH with positive entropy and vice-versa. Much work is 
required for the construction of the formulation which describes such 
phase transitions (if they really occur).

In any case, the BH states with zero (or negative) entropy call to 
the further investigation as many previous claims on the properties 
of BH entropy may not be  true. Just to give an example,
there is a conjecture in ref.\cite{bousso} that the entropy  of deSitter 
space corresponds to the upper bound for the entropy of the 
asymptotically deSitter space. It is clear that above conjecture maybe
 true only in the Einstein gravity.
In higher derivative theories, say, in the case of zero (or negative) 
SdS BH entropy this bound cannot be applied. 
Hence, the complete understanding of gravitational entropy for higher 
derivative theories remains the subject of future research.

An important question to be raised has to do  whether the type of
gravities and the parameter range discussed in this paper can be realized
as an M-theory and (or) string compactification. In particular, it is well 
known that de Sitter space 
(or de Sitter supergravity) is hard to realize as vacuum state in 
this context. One possible
candidate  is a compactification of  Type IIB* supergravity \cite{ds1}.
However, since Type IIB* supergravity is obtained 
from  Type IIB by performing
a T-duality along the time direction, that introduces negative kinetic
energy terms (ghosts) for field strengths F in the
new Type IIB theory. Another (not very realistic) possibility could be 
related with gauged supergravity where de Sitter state (at least as maximum)
can be realized.

The  parameter ranges for the higher derivative terms employed in this
paper may also require inclusion of other types of higher derivative
terms. In principle such higher derivative terms 
are calculable in string  and M-theory   and a further compactification 
on Einstein-Sasaki-type spaces would  yield  a lower dimensional gravity
with a (negative) cosmological constant.  Thus in string theory the
magnitude of the higher derivative terms and the
magnitude of the cosmological constant are in principle calculable
parameters in terms of the string coupling 
and the parameters of the compactified space.  It could  turn out
that the domain of parameters discussed in this paper would require the
inclusion of other higher derivative terms in string theory  which would 
in turn require a modification of our analysis and is deferred for a
future study.

\section*{Acknoweledgements} 
MC and SDO would like to thank the organizers of First Mexican 
Meeting on Math. and Exp. Physics, where this work was initiated, for 
hospitality. SDO would like to thank S. Hawking for kind 
hospitality in DAMTP where this work was completed.
We are grateful to G. Gibbons, J. Lidsey and A.Naqvi for helpful 
discussions. 
The work by SN is supported in part by the Ministry of Education, 
Science, Sports and Culture of Japan under the grant n. 13135208.
This research is supported in part by DOE grant DE-FG02-95ER40893, 
NATO Linkage grant 976951 and the Class of 1965 Endowed Term Chair.

\end{document}